\documentclass[pra,twocolumn]{revtex4}  

\usepackage{xcolor}
\usepackage{amssymb}
\usepackage{amsmath}
\usepackage{graphicx}
\usepackage{graphicx}
\graphicspath{{/Users/joanagusti/Dropbox/LaTex/SuperQuLan/Paper/v10/Figures}}
\usepackage{float}
\usepackage{braket}
\usepackage{bbm}
\usepackage{color}
\usepackage{comment}
\usepackage{gensymb}
\usepackage{gensymb}
\usepackage{lipsum}
\definecolor{blue}{rgb}{0,0,1}
\definecolor{grey}{rgb}{0.6,0.6,0.6}

\usepackage[colorlinks=true,citecolor=blue,linkcolor=blue,urlcolor=blue]{hyperref}

\begin{document}

\title{Long-distance distribution of qubit-qubit entanglement using Gaussian-correlated photonic beams}

\author{J. Agust{\'i}­$^1$}
\author{Y. Minoguchi$^1$}
\author{J. M. Fink$^2$}
\author{P. Rabl$^1$}
\affiliation{$^1$ Vienna Center for Quantum Science and Technology, Atominstitut, TU Wien, 1020 Vienna, Austria\\
$^2$ Institute of Science and Technology Austria, am Campus 1, 3400 Klosterneuburg, Austria}

\date{\today}
\begin{abstract}
We investigate the deterministic generation and distribution of entanglement in large quantum networks by driving distant qubits with the output fields of a non-degenerate parametric amplifier. In this setting, the amplifier produces a continuous Gaussian two-mode squeezed state, which acts as a quantum-correlated reservoir for the qubits and relaxes them into a highly entangled steady state. Here we are interested in the maximal amount of entanglement and the optimal entanglement generation rates that can be achieved with this scheme under realistic conditions taking, in particular, the finite amplifier bandwidth, waveguide losses and propagation delays into account. By combining exact numerical simulations of the full network with approximate analytic results, we predict the optimal working point for the amplifier and the corresponding qubit-qubit entanglement under various conditions. Our findings show that this passive conversion of Gaussian- into discrete-variable entanglement offers a robust and experimentally very attractive approach for operating large optical, microwave or hybrid quantum networks, for which efficient parametric amplifiers are currently developed.   
\end{abstract}

\maketitle

\section{Introduction}
\label{sec:Introduction}
The distribution of entanglement between separated nodes of small- and large-scale quantum networks is a fundamental task for many quantum communication and quantum information processing applications~\cite{Nielsen2000,Kimble2008,Wehner2018,Kim2014,Lukin2007,Bianchi2020}. Once established, entanglement can be locally purified~\cite{Bennett1996,Duer1999,Macchiavello1999,Massar1999,Cleland2022} and used for quantum teleportation and remote gate operation protocols that then require classical communication only~\cite{Chuang1999}. The availability of a large number of highly entangled qubit pairs, shared among different nodes, is thus a universal and in practice one of the most essential resources for quantum network applications.

Existing protocols for distributing entanglement in realistic systems are mostly based on one of the following two strategies~\cite{Northup2014}: A first approach is to generate a pair of entangled qubits locally and then transfer one of the states to the distant node, for example, through a controlled emission and reabsorption of optical or microwave photons~\cite{Cirac1997, Kurpiers2018,Schoelkopf2018,Wallraff2020,Cleland2021}. This protocol is fully deterministic but requires a sufficiently high level of (synchronous) control on both nodes of the network.  A second strategy is to generate only weakly correlated qubit-photon pairs from which highly entangled qubit states can be distilled through measurements and postselection~\cite{Duan2001,Barrett2005,Matsukevich2006,Moehring2007,Hensen2015,Riedinger2018}. This approach has the advantage that it requires only little local control and that it is intrinsically robust with respect to losses. It is, however, only probabilistic and for many implementations~\cite{Matsukevich2006,Moehring2007,Hensen2015,Riedinger2018} the scalability of this approach is limited by the low success probabilities, which also decrease exponentially with the number of qubit pairs that must be entangled simultaneously in this way.


\begin{figure}[b]
	\centering
	\includegraphics[width=\columnwidth]{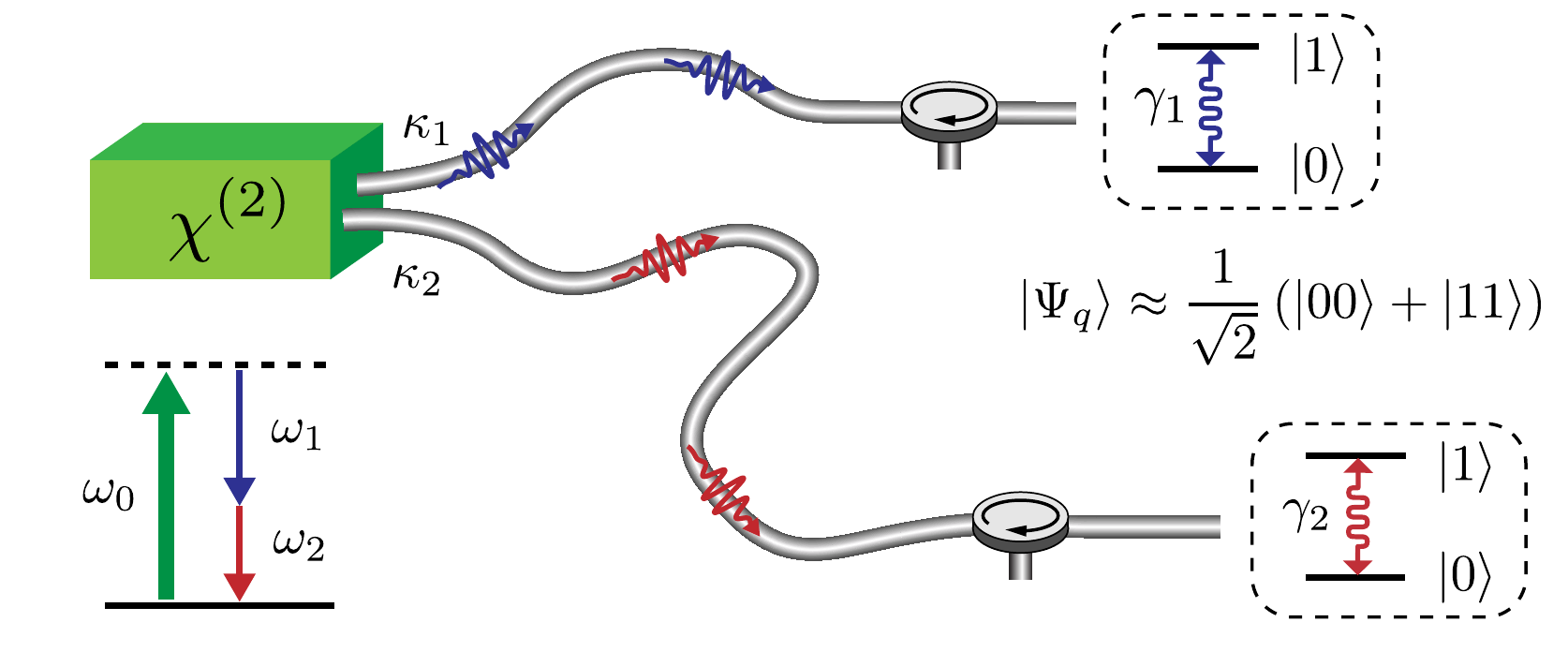}
	\caption{Setup. A non-degenerate parametric amplifier is used to produce two Gaussian-correlated beams of photons, which drive two qubits located in separate notes of the network into an entangled state. See text for more details. }
	\label{fig:Fig1Setup}
\end{figure}

A complimentary third strategy for distributing entanglement is to drive distant qubits with continuous beams of quantum-correlated photons, as shown in Fig.~\ref{fig:Fig1Setup}. In the simplest case, entangled beams of optical and/or microwave fields can be generated in a parametric down-conversion process \cite{Textbooks_CarmichaelOptics}, which produces as an output a propagating two-mode squeezed state. This process only requires a weak intrinsic nonlinearity as it occurs, for example, in nonlinear optical crystals \cite{Zeilinger1995,Wong2006} or in driven Josephson junctions in the microwave regime \cite{Eichler2011,Menzel2012,Huard2012,Devoret2016,Winkel2020,Peugeot2021}. 
Therefore, parametric down-conversion is currently the most common method to generate entangled pairs of optical photons \cite{Walmsley2011,Couteau2018},
however, usually in a probabilistic and postselected manner. In contrast, here we are interested in the regime where the parametric amplifier is strongly pumped, such that the output fields contain many photons on average. By using these correlated photons to drive two spatially separated qubits, the Gaussian entanglement can be mapped onto an entangled qubit state. This scheme has the obvious benefit that it only relies on an externally pumped $\chi^{(2)}$-nonlinearity for generating the entanglement, which is typically much easier to realize than strong few-photon interactions or high-fidelity qubit-qubit gates.  At the same time, this approach does not rely on postselection and can be used to distribute entanglement deterministically.

The basic idea of using two-mode squeezed states of light to drive qubits into highly entangled states has been originally discussed in a paper by Kraus and Cirac~\cite{Kraus2004}, and since then several related schemes for engineering correlated reservoirs for qubits have been described~\cite{Paternostro2004,Serafini2006,Muschik2011,Felicetti2014,Blais2017,Mirrahimi20018,You2018,Govia2021}. Importantly, these works already show that under ideal conditions a maximally entangled qubit state can be prepared in this way. In view of the vastly different nature of entangled continuous-variable states and entangled discrete-variable states, this result is a priori not apparent. However, all these predictions rely on the crucial assumption that the correlated photon source can be treated as an effective Markovian reservoir, which ignores the finite bandwidth of the parametric amplifier, propagation delays and other effects that are relevant for practical networking applications. Therefore, a systematic analysis of the performance of this approach under realistic conditions, for example, in terms of speed, robustness, or the fidelity of the produced entangled states, is still wanting.

In this paper, we study the entanglement properties of two distant qubits that are driven by the output of a realistic two-mode parametric amplifier. In this analysis we go beyond the usual Markov approximation and use both exact numerical simulations as well as approximate analytic methods to evaluate the dynamics and steady states of the qubits, taking the finite bandwidth of the amplifier and other relevant sources of imperfections fully into account. This analysis allows us to predict optimal operation points for the parametric amplifier and assess the maximal fidelities and entanglement generation rates that can be reached with this scheme under given experimental constraints. These findings do not rely on any specific implementation and can thus be applied for the optimization of remote entanglement distribution schemes in various optical, microwave or hybrid~\cite{Xiang2013,Kurizki2015} quantum networks.

The remainder of the paper is organized as follows: In Sec.~\ref{sec:Setup} we introduce the setup and the individual network components. In Sec.~\ref{sec:Markov} we first briefly review the generation of qubit-qubit entanglement in the infinite-bandwidth limit and discuss the relationship between the achievable amount of entanglement and the basic characteristics of the squeezed reservoir.   In Sec.~\ref{sec:NonMarkov} we go beyond the Markov approximation and use numerical and approximate analytic results to evaluate the steady-state entanglement in realistic networks. Finally, in Sec.~\ref{sec:LongDistance} and Sec.~\ref{sec:Entanglement} we discuss the performance of the scheme in asymmetric networks as well as the optimal entanglement distribution rates in a pulsed operation mode. Finally, in Sec.~\ref{sec:Conclusions} we summarize our main findings and place them into context with respect to the current experimental state of the art.

\section{Setup}
\label{sec:Setup}
We consider the generic setting depicted in Fig.~\ref{fig:Fig1Setup}, where two distant qubits with frequencies $\omega_{q1}$ and $\omega_{q2}$ are driven by the  output fields of a non-degenerate parametric amplifier. The amplifier consists of two distinct bosonic modes with frequencies $\omega_1$ and $\omega_2$ and annihilation operators $a_1$ and $a_2$, respectively. These modes are driven into a correlated two-mode squeezed state via an externally pumped $\chi^{(2)}$-process, and each of them is connected to one of the qubits through a unidirectional channel.  In practice, such a scenario can be realized with the help of circulators, directional couplers or chiral waveguides (see, for example, Refs.~\cite{Cirac1997,Stannigel2014,Lodahl2017}). The following analysis is kept deliberately general and applies to implementations in the optical and microwave domain, as well as mixed scenarios, where, for example, correlated pairs of optical and microwave photons are generated via the electro-optical effect~\cite{Rueda2019,Krastanov2021}. However, throughout our analysis we will assume a matching of the corresponding qubit and photon frequencies, i.e., $\omega_{qi}\approx \omega_i$, and that all parts of the network are sufficiently cold such that thermal excitations can be neglected \cite{Huard2015}.

\subsection{Qubits}
\label{subsec:Qubits}
We model the qubits as simple two-level systems with a ground state $|0\rangle$ and an excited state $|1\rangle$. The qubits decay into the waveguide with rates $\gamma_i$ and undergo dephasing with rate $\gamma_\phi$. In the absence of the amplifier and by changing into a rotating frame with respect to the qubit frequencies $\omega_{q1}$ and $\omega_{q2}$, the resulting dynamics for the qubit density operator $\rho_q$ is given by the master equation  
\begin{equation}
\dot \rho_q =  \sum_{i=1,2} \gamma_i \mathcal{D}[\sigma^-_i,\sigma_i^+] \rho_q + \frac{\gamma_\phi}{2} \mathcal{D}[\sigma^z_i,\sigma^z_i] \rho_q \equiv \mathcal{L}_{q}\rho_q.
\end{equation}
Here, $\sigma^-=(\sigma^+)^\dag=|0\rangle\langle 1|$ and $\sigma^z=|1\rangle\langle 1| - |0\rangle\langle 0|$ and  we have introduced the short notation 
\begin{equation}
\mathcal{D}[A,B] \rho = A\rho B- \frac{1}{2} \left(B A \rho + \rho B A \right)
\end{equation} 
for the Liouville superoperators describing incoherent decay and dephasing.

\subsection{Parametric amplifier}
\label{subsec:Param}
Parametric amplification relies on a $\chi^{(2)}$-type process between three bosonic modes $(\hbar=1)$ \cite{Textbooks_CarmichaelOptics},
\begin{equation}
H_{\rm \chi} = i \chi( a_{1}^{\dagger}a_{2}^{\dagger}a_{0}- a_{1}a_{2}a_{0}^{\dagger}),
\end{equation}
where the nonlinearity $\chi$ is small, but one of the modes, $a_0$, is strongly pumped. Under this assumption, the pumped mode can be treated as a classical field, $a_0\rightarrow \alpha_0 \in \mathbbm{C}$, and $H_{\rm \chi}$ reduces to a two-mode squeezing interaction.  This process becomes most effective when the resonance condition $\omega_1+\omega_2\approx \omega_0$ between the three modes is fulfilled (see Fig.~\ref{fig:Fig1Setup}) and is frequently employed in nonlinear optical crystals to produce entangled photon pairs. Similar interactions also occur in superconducting circuits, where driven Josephson junctions are used to generate two-mode squeezed microwave beams, and in many other devices. In the following, we fix the pump frequency to a value of $\omega_0=\omega_{q1}+\omega_{q2}$, in order to maximize the resulting qubit-qubit correlations discussed below. The residual photon-qubit detunings are denoted by $\Delta_i=\omega_i-\omega_{qi}$.

The parametrically generated pairs of photons in modes $a_1$ and $a_2$ decay into the waveguides with rates $\kappa_1$ and $\kappa_2$, respectively.  
By changing into a rotating frame with respect to $\omega_{q1}$ and $\omega_{q2}$, the full dynamics of the two-mode photonic state $\rho_p$ is described by the master equation
\begin{equation}\label{eq:MEParam}
\dot \rho_p  = -i\left[H_p, \rho_p\right] + \sum_{i=1,2} \kappa_i \mathcal{D}[a_i,a_i^\dag] \rho_p\equiv \mathcal{L}_{p}\rho_p,
\end{equation}
where
\begin{equation}\label{eq:HamiltonianParam}
H_p = \sum_{i=1,2} \Delta_i a_i^\dag a_i  +  i \frac{\sqrt{\kappa_1\kappa_2} \epsilon}{2}   \left( a_{1}^{\dagger}a_{2}^{\dagger}  - a_{1}a_{2} \right)
\end{equation} 
 and $\epsilon\sim \chi |\alpha_0|$ is the dimensionless pump parameter, which we can assume to be real. For $\Delta_i=0$, the value of $\epsilon=1$ marks the onset of the parametric instability, beyond which our linearized description of the amplifier is no longer valid.  For the remainder of this work, we focus mainly on the resonant case and restrict the pumping parameter to $\epsilon \in [0,1)$.

\subsection{Cascaded photon-qubit interactions}
\label{subsec:Casc} 
In the considered cascaded setting, photons emitted by the parametric amplifier will drive the qubits located further along the waveguides, while there is no backaction of the qubits on the photonic modes. By assuming a waveguide with a broad and linear dispersion relation, this scenario can be modelled in terms of the cascaded master equation~\cite{Carmichael1992,Gardiner1992} 
\begin{equation}\label{eq:CascadedME}
	\dot \rho=\left(\mathcal{L}_{q}+\mathcal{L}_{p}+\mathcal{L}_{\rm cas}\right)\rho,
\end{equation}
where $\rho$ is the density operator of the whole system, including both the qubits and the photonic degrees of freedom. The last term in Eq.~\eqref{eq:CascadedME}, 
\begin{equation}
\mathcal{L}_{\rm cas}\rho =\sum_{i=1,2}\sqrt{\eta \gamma_{i}\kappa_{i}}\left([a_{i}\rho,\sigma_{i}^{+}]+[\sigma^-_{i},\rho a_{i}^{\dagger}]\right),
\end{equation} 
accounts for the cascaded, i.e. unidirectional coupling between the photons and the qubits. Here we have introduced the additional phenomenological parameter $\eta \in [0,1]$. 
A value of  $\eta<1$ can either be used to model losses along the waveguide, in which case $(1-\eta)$ is the probability that a propagating photon is lost between the amplifier and the qubit, or additional decay channels for the qubits other than the emission into the waveguide.

Note that in writing Eq.~\eqref{eq:CascadedME} we have already reabsorbed all propagation phases into a redefinition of the qubit states. For now, we have also neglected the finite propagation times $\tau_i=d_i/c_i$, which it takes for photons with group velocity $c_i$ to travel the distance $d_i$ between the amplifier and the $i$-th qubit. This is usually a valid assumption for small on-chip networks, but propagation delays can become very relevant when discussing entanglement distribution in larger networks, in particular when $d_1\neq d_2$. This issue will be addressed in more detail in Sec.~\ref{sec:LongDistance} below.

\subsection{Reduced qubit state}
Starting from the cascaded master equation of the full network, Eq.~\eqref{eq:CascadedME}, our main goal in the following is to study the dynamics and stationary states of the reduced qubit state, $\rho_q(t)={\rm Tr}_p\{\rho(t)\}$, as a function of the pumping strength $\epsilon$ and other system parameters. From the structure of the superoperator $\mathcal{L}_{\rm cas}$ it is straightforward to show that the cascaded coupling doesn't affect the dynamics of the reduced state of the amplifier, i.e., $\dot \rho_p= {\rm Tr} _q\{ \mathcal{L}_{\rm cas} \rho \}=0$. However, this doesn't imply that the system can be factorized into a photonic and a qubit part. In contrast, it can be explicitly shown that in such cascaded systems a large amount of entanglement between the individual subsystems can emerge~\cite{Stannigel2014}. Therefore, in general, it is necessary to solve the full master equation in order to obtain accurate predictions for $\rho_q$.

\section{Qubits coupled to a two-mode squeezed reservoir} 
\label{sec:Markov}
As a starting point, it is instructive to revisit the idealized limit of a broadband amplifier, $\kappa_i\rightarrow \infty$, where the dynamics of the amplifier modes is much faster than the qubit evolution and can be adiabatically eliminated. In this case, it is indeed possible to obtain a reduced master equation for the qubit state  $\rho_q$ only, which reads
\begin{equation}\label{eq:EffectiveME}
\begin{split}
	\dot \rho_q&=\mathcal{L}_q\rho_q +  \sum_{i=1,2} \sum_{s=\pm}     \gamma_i   N_{i}\mathcal{D}[\sigma_i^s,\sigma_i^{-s }]\rho_q \\
	&-\sqrt{\gamma_1\gamma_2} \left\{ M\left( \mathcal{D}[\sigma_1^+,\sigma_2^+]\rho_q+ \mathcal{D}[\sigma_2^+,\sigma_1^+]\rho_q\right)+ \rm H.c.\right\}.
\end{split}
\end{equation}
Here the occupation numbers $N_i = 2\eta \mathrm{Re} \{ I^{\rm}_{a_i^\dag a_i}(0)\} $ and the squeezing term $M = \eta [I^{\rm}_{a_1 a_2}(0)+I^{\rm}_{a_2 a_1}(0)] $ are determined by the steady-state correlation spectra of the amplifier modes, $I_{A_iB_j}(\omega)=\sqrt{\kappa_i \kappa_j}    \int_{0}^{\infty}\mathrm{d}t \, \braket{A_i(t)B_j(0)} e^{-i\omega t}$, evaluated at frequency $\omega=0$. A detailed derivation of this master equation and explicit expressions for the two-time correlation functions are given in Appendix~\ref{sec:AppendixME} and Appendix~\ref{sec:AppendixCorrelations}, respectively.

In the symmetric case, $\kappa_i=\kappa$, and for $\Delta_i=0$ we obtain the simple results 
 \begin{eqnarray}
	2\mathrm{Re} \{I_{a_i^\dag a_i}(\omega)\}=\epsilon\left[\Gamma_{-}(\omega)-\Gamma_{+}(\omega)\right],
\end{eqnarray}
and
 \begin{eqnarray}
	I_{a_1 a_2}(\omega)+I_{a_2 a_1}(-\omega)=
		\epsilon\left[\Gamma_{-}(\omega)+\Gamma_{+}(\omega)\right],
\end{eqnarray}
where
\begin{equation}
	\Gamma_{\pm}(\omega)=\frac{\kappa^{2}}{\kappa^{2}(1\pm\epsilon)^{2}+4\omega^{2}}
\end{equation}
are Lorentzian functions. The parameters  $N=N_i$ and $M$ in Eq.~\eqref{eq:EffectiveME} are then given by
\begin{eqnarray}
	N&=&\epsilon \eta \left(\frac{1}{\left(1-\epsilon\right)^{2}}-\frac{1}{\left(1+\epsilon\right)^{2}}\right), \label{eq:MarkovN}\\
		M&=&\epsilon \eta \left(\frac{1}{\left(1 -\epsilon\right)^{2}}+\frac{1}{\left(1 +\epsilon\right)^{2}}\right).\label{eq:MarkovM}
\end{eqnarray}

\subsection{Squeezing and purity}\label{subsec:SqueezingPurity}
To obtain additional intuition, we focus on the symmetric configuration $\kappa_i=\kappa$, where we can reinterpret $N =\langle a^\dag_i a_i\rangle_{\rho_{\rm eff}}$ and  $M =\langle a_1 a_2\rangle_{\rho_{\rm eff}} $ as expectation values of an effective two-mode squeezed state 
\begin{equation}
\rho_{\rm eff}= S(r_{\rm eff}) \rho_{\rm th}( \overline{n}_{\rm eff}) S^\dag(r_{\rm eff}).
\end{equation}
Here, $\rho_{\rm th}( \overline{n}_{\rm eff})$ is a thermal two-mode state with occupation number $ \overline{n}_{\rm eff}$ and 
\begin{equation}
S(r)=e^{r (a_1^\dag  a_2^\dag - a_1 a_2)} 
\end{equation}
is the two-mode squeezing operator. Instead of the thermal occupation number, we can also work with the purity of the effective state, $\mu_{\rm eff}=1/(2\overline{n}_{\rm eff}+1)$, in which case we obtain the relations~\cite{Serafini04_TwoModeGaussianNoise}
\begin{eqnarray}\label{eq:sq_eff}
	r_{\rm eff}&=&\frac{1}{2}\tanh^{-1}{\left[\frac{2|M|}{2N+1}\right]}, \\
	\mu_{\rm eff}&=&\frac{1}{\sqrt{(2N+1)^{2}-4|M|^{2}}}. \label{eq:mu_eff}
\end{eqnarray} 
From the expressions given in Eq.~\eqref{eq:MarkovN}  and Eq.~\eqref{eq:MarkovM} we find that $ |M|^{2}= N(N+\eta)$ and therefore $\mu_{\rm eff}=1$ for a lossless channel. This implies that in the infinite-bandwidth limit,  Eq.~\eqref{eq:EffectiveME} describes two qubits that are coupled to a two-mode squeezed zero-temperature reservoir with a squeezing parameter 
\begin{equation}
	r_{\rm eff}(\epsilon)=2\tanh^{-1}(\epsilon),
\end{equation} 
which becomes arbitrarily large when $\epsilon\rightarrow 1$. However, this is a consequence of the Markov approximation, where the environment is only probed at a single frequency. Below we will show that this is no longer true when a nonvanishing ratio $\kappa/\gamma$ is taken into account, in which case averaging over a finite frequency window $\Delta\omega\neq 0$ translates into an impure effective state with reduced squeezing. The same is already true in the Markovian limit in the presence of transmission losses, $\eta<1$.

%

\subsection{Steady state entanglement}\label{sec:SteadyState}
From the reduced master equation in Eq.~\eqref{eq:EffectiveME} we can derive the steady state of the two qubits, which we express in general as  
\begin{equation}
\rho_q^0= \sum_{\boldsymbol{s},\boldsymbol{s}^\prime} \rho^0_{\boldsymbol{s},\boldsymbol{s}^\prime} |\boldsymbol{s}\rangle\langle \boldsymbol{s}^\prime|,
\end{equation} 
\noindent where $\boldsymbol{s}=(s_1,s_2)$ with $s_i=0,1$ labels the two-qubit states.
For a symmetric parametric amplifier, $\kappa_1=\kappa_2$, and for symmetric qubits, $\gamma_1=\gamma_2$, there are only six nonvanishing matrix elements,
\begin{eqnarray*}
	\Lambda \rho^0_{00,00}&=&(1+N)^{2}(2\Gamma_{\phi}+1+2N)-|M|^{2}(3+2N),\\
	\Lambda\rho^0_{10,10}&=&2\Gamma_{\phi}N(N+1)+(2N+1)(N(N+1)-|M|^2),\\
	\Lambda \rho^0_{11,11}&=&N^2(1+2\Gamma_{\phi}+2N)+|M|^2(1-2N),\\
	\Lambda \rho^0_{11,00}&=&M,
\end{eqnarray*}
\noindent and $\rho^0_{01,01} =\rho^0_{10,10}$ and $\rho^0_{00,11} =(\rho^{0}_{11,00})^*$. Here we introduced the normalization constant $\Lambda=(1+2N)[1+2\Gamma_{\phi}+4(N(N+1+\Gamma_{\phi})-|M|^2))]$ and the normalized dephasing rate $\Gamma_{\phi}=\gamma_{\phi}/\gamma$. In the Markov limit and assuming also otherwise ideal conditions, $\gamma_{\phi}=0$ and $\eta=1$, we have $|M|^{2}=N(N+1)$, $\rho^0_{10,10}=0$ and $|\rho^0_{11,00}|^2= \rho^0_{11,11} \rho^0_{00,00}$. This means that the steady state is the pure state $\rho_q^0=\ket{\Psi_q}\bra{\Psi_q}$,
where \cite{Cirac1997}
\begin{equation}\label{eq:Psiq}
	\ket{\Psi_q}=\sqrt{\frac{N+1}{1+2N}}\ket{00}+e^{i\theta}\sqrt{\frac{N}{1+2N}}\ket{11},
\end{equation}
and we defined $M=|M|e^{i\theta}$. On resonance, $\Delta_i=0$, we obtain $\theta=0$ and the steady state approaches the maximally entangled triplet state $\ket{\Phi^{+}}=\frac{1}{\sqrt{2}}(\ket{00}+\ket{11})$ for $N\gg 1$. 

Before we proceed, let us provide some additional insights about the emergence of such a pure entangled steady state, by considering the coupling of the qubits to two isolated modes $a_1$ and $a_2$ via a Jaynes-Cummings interaction of the form
\begin{equation}
	H_{\rm int} \sim i\left(\sigma_1^-a_1^\dagger - \sigma_1^+a_1 +\sigma_2^-a_2^\dagger - \sigma_2^+a_2\right).
\end{equation}
An ideal two-mode squeezed state of those two modes can be written as 
\begin{equation}
	\ket{\Psi_{\rm TMS}}=\frac{1}{\sqrt{1-x^2}} \sum_{n=0}^\infty x^n |n\rangle_1|n\rangle_2,
\end{equation}
where we can set $x=\sqrt{N/(N+1)}$ to match the convention from above. This expression shows that the number of photons in the two modes are perfectly correlated, suggesting that the qubits are only excited and de-excited pairwise. However, this argument is too naive, since, for example, the action of $H_{\rm int}$ on the state $|00\rangle\ket{\Psi_{\rm TMS}}$ would also generate singly-excited states $\sim|10\rangle, |01\rangle$. To explain the existence of the steady state given in Eq.~\eqref{eq:Psiq} it is thus important to take into account the coherence between the $ |n\rangle_1|n\rangle_2$ components, which leads to  the following relations
\begin{eqnarray}
a_1 \ket{\Psi_{\rm TMS}} &=&x a_2^\dag \ket{\Psi_{\rm TMS}},\\
a_2 \ket{\Psi_{\rm TMS}} &=&x a_1^\dag \ket{\Psi_{\rm TMS}}.
\end{eqnarray} 
Equivalently, there exists a unique dark state of the interaction,
\begin{equation}
	H_{\rm int}(|00\rangle+x|11\rangle)\ket{\Psi_{\rm TMS}}=0.
\end{equation}
Therefore, once the system reaches the state $\ket{\Psi_q}$ in Eq.~\eqref{eq:Psiq}, the emission of a photon by one qubit interferes destructively with the absorption of a photon in the other mode. When applied to the original setting, this argument shows that the absence of any components $\sim |01\rangle$ or $\sim |10\rangle$ in $|\Psi_q\rangle$ is a consequence of a nonlocal interference effect between two well-separated, but correlated parts of the network.

\subsection{Non-ideal squeezed reservoirs}
To investigate the performance of this entanglement distribution scheme also under non-ideal conditions, for example, when $|M|^{2}<N(N+1)$ or when a reduced master equation for $\rho_q$ is not available, we consider the fidelity
\begin{equation}\label{eq:Fidelity}
	\mathcal{F}=|\braket{\Phi^{+}|\rho^0_{q}|\Phi^{+}}|^{2}\leq \frac{(1+\epsilon)^{4}}{2(1+6\epsilon^{2}+\epsilon^{4})}
\end{equation}
as a measure of how close the qubit state approaches the maximally entangled triplet state. In addition, we can use the concurrence of the reduced state, $\mathcal{C}(\rho_q^0)$ \cite{Horodeki2009}, to quantify directly the amount of qubit-qubit entanglement.  Note that the upper bound in Eq.~\eqref{eq:Fidelity} derived for the ideal Markovian limit shows that already for moderate driving strengths, $\epsilon\gtrsim 0.5$, fidelities of about $\mathcal{F}\approx 0.99$ can be reached.

\begin{figure}
	\centering
	\includegraphics[width=\columnwidth]{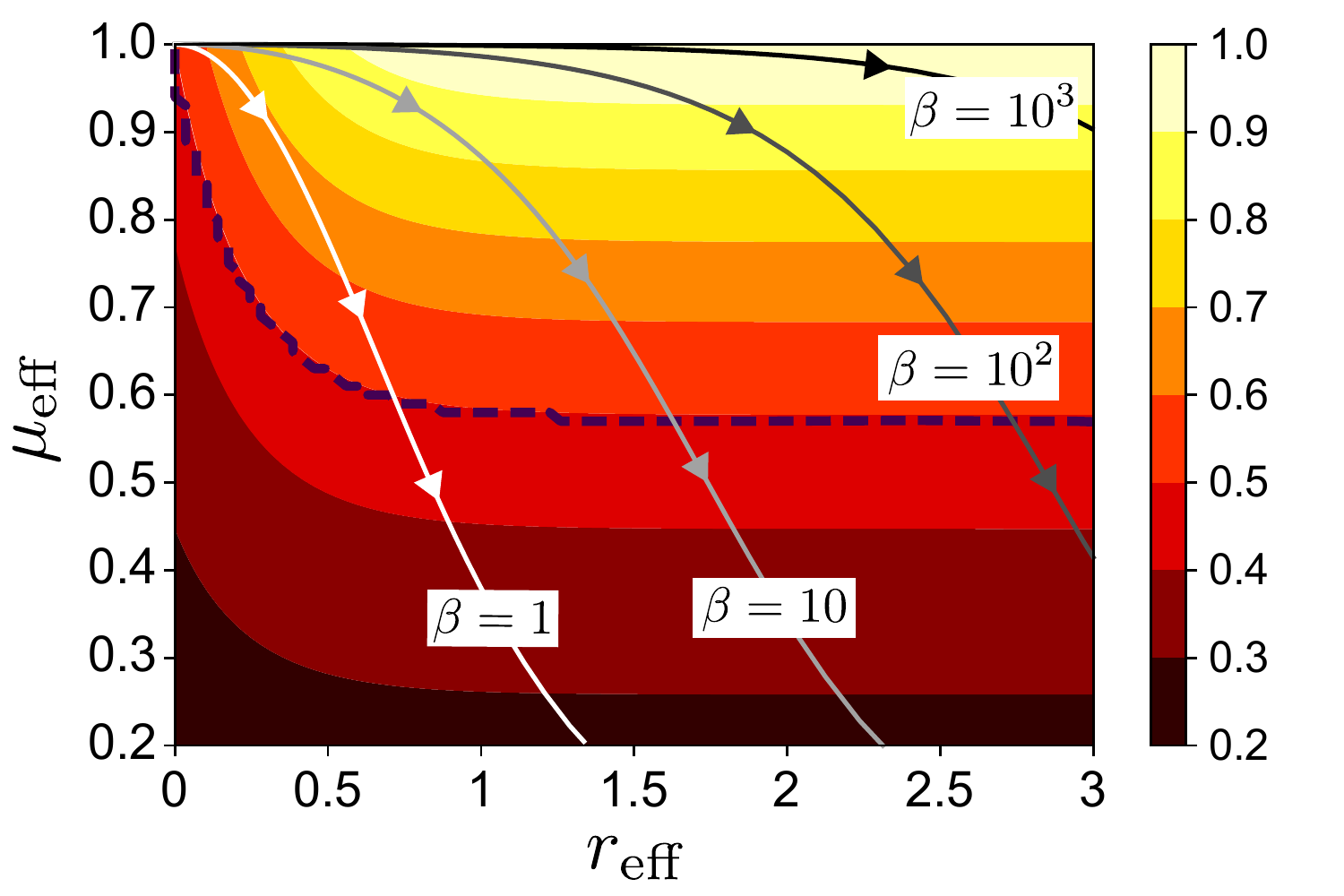}
	\caption{Contour plot of the steady state fidelity $\mathcal{F}$ defined in Eq.~\eqref{eq:Fidelity} as a function of the effective squeezing parameter $r_{\rm eff}$ and the effective purity $\mu_{\rm eff}$. The solid lines indicate the path in this parameter space that one obtains by increasing $\epsilon$ from 0 to 1 for different values of $\beta=\kappa/\gamma$ and $\Gamma_\phi=0$ and $\eta=1$ (see discussion in Sec.~\ref{sec:NonMarkov}). The dashed line marks the boundary of vanishing concurrence $\mathcal{C}=0$, above which the reduced qubit state is entangled. }
	\label{fig:MarkovContour}
\end{figure}

In Fig. \ref{fig:MarkovContour} we show a plot of the steady state fidelity $\mathcal{F}$ for the steady state of master equation~\eqref{eq:EffectiveME}, under the assumption that $\Gamma_\phi=0$, but allowing for arbitrary values of $N_i=N$ and $|M|^2\leq N(N+1)$. For later convenience, these parameters are in turn expressed in terms of the effective squeezing parameter $r_{\rm eff}$ and the effective purity $\mu_{\rm eff}$, as defined in Eq.~\eqref{eq:sq_eff} and Eq.~\eqref{eq:mu_eff}. We see that for $\mu_{\rm eff}\simeq 1$, the fidelity approaches unity as the squeezing parameter is increased, consistent with the bound stated in Eq.~\eqref{eq:Fidelity}. 
For values of $r_{\rm eff}\gtrsim 1$, the fidelity becomes almost independent of the squeezing parameter, but decreases as $\mathcal{F}\simeq \frac{1}{4}(1+3\mu^2_{\rm eff})$ for impure reservoirs. Entanglement is present as long as $\mathcal{F}> 0.5$, which corresponds to a minimal purity of about  $\mu_{\rm eff}\simeq1/\sqrt{3}\simeq 0.57$. This result shows that improving the purity of the effective photonic bath will be most relevant for this entanglement distribution scheme.

\section{Entanglement distribution in realistic networks} 
\label{sec:NonMarkov}
Let us now address the more realistic scenario, where finite waveguide losses, decoherence of the qubits and, in particular, the finite bandwidth of the parametric amplifier are taken into account. In this case, it is in general no longer possible to eliminate the photon modes and the full cascaded master equation in Eq.~\eqref{eq:CascadedME} must be solved numerically. In view of the large Hilbert space required to represent the two-mode squeezed state, these simulations become very demanding when approaching the parametric instability, $\epsilon=1$. Thus, in the following, we restrict all numerical simulations to values of $\epsilon\leq 0.8$, where a convergence of the results can still be ensured by truncating the Hilbert space of each photon mode to $n_{\rm trunc}\lesssim 25$ states and by performing finite size scaling analysis (see Appendix~\ref{app:Truncation}). 
In addition, we introduce a filtered-mode approximation (FMA) in order to obtain also an analytic dependence of the fidelity on all the relevant system parameters.       

\subsection{Filtered mode approximation} 
In the Markov limit studied in the previous section, the state of the qubits obeys a master equation, where the characteristic bath parameters $N_{i}$ and $M$ are determined by the output fields of the amplifier at a single frequency, $\omega=0$. To go beyond this approximation, we must take into account that the qubits will be affected by photons within a finite region of the spectra $S_{a_i^\dag a_j}(\omega)$ and $S_{a_i a_j}(\omega)$ that cannot be associated with a pure squeezed state. The relevant bandwidth of frequencies will be determined by the dynamics of the qubits themselves and will be roughly given by the decay rates $\gamma_i$. Based on this intuition, we introduce the two filtered modes
\begin{equation} \label{eq:filtered_modes}
	a_{f,i}(t)=\sqrt{\gamma_{i} \kappa_i\eta}\int_{-\infty}^{t} {\rm d}s \,e^{-\gamma_{i}(t-s)/2} a_{i}(s-\tau_i),
\end{equation}
where, for a later generalization, we have already included the propagation delays $\tau_i$. These modes represent the output of the two-mode amplifier, but delayed by $\tau_i$ and filtered by the response of the qubits. 

We can now use these filtered modes to define an adjusted set of parameters for the qubit master equation in Eq.~\eqref{eq:EffectiveME}, 
\begin{equation}
N_i=\langle a^\dag_{f,i} a_{f,i}\rangle, \qquad M=\langle   \mathcal{T} a_{f,1} a_{f,2}\rangle,
\end{equation} 
where $ \mathcal{T}$ denotes the time-ordering operator applied to the amplifier modes $a_1$ and $a_2$. These parameters include the characteristic timescales of the qubits and of the photons on an equal footing. Specifically, we obtain
\begin{equation}\label{eq:Eff_mode_N}
N_i=2 \eta \gamma_i   \int_{-\infty}^{\infty} \frac{\mathrm{d}\omega}{2\pi}  \frac{I_{a_i^\dag a_i}(\omega)}{\gamma_i^2/4+\omega^2} 
\end{equation}
for the occupation numbers and 
\begin{equation} \label{eq:Eff_mode_M}
	\begin{split}
	M= \sqrt{\gamma_1\gamma_2}\eta \int_{-\infty}^{\infty}\frac{\mathrm{d}\omega}{2\pi}  \frac{\left[ I_{a_1a_2}(\omega)+I_{a_2a_1}(-\omega)\right] e^{i\omega (\tau_2-\tau_1)}}{(\gamma_1/2+i\omega)(\gamma_2/2-i\omega)}
	\end{split}
\end{equation}
for the correlation parameter. These expressions explicitly show how the effective photonic reservoir seen by the qubits depends on the amplifier correlations within finite frequency windows set by the decay rates $\gamma_i$.  The Markovian limit discussed in Sec.~\ref{sec:Markov} is recovered in the limit $\gamma_i\rightarrow 0$ and $\tau_i\rightarrow 0$.

\subsection{Effective squeezing parameters for non-ideal amplifiers} 
While the derivation of a master equation in terms of the filtered modes is only an approximation, it becomes exact in the regime of low qubit excitations, i.e. for small values of $\epsilon$ (see Appendix~\ref{app:TimeDelayQLE}). Even for moderate and large driving strengths it still results in a considerable improvement over the conventional master equation discussed in Sec.~\ref{sec:Markov}. In particular, the FMA allows us to account for the effects of a finite amplifier bandwidth. For example, by setting $\kappa_i=\kappa$, $\gamma_i=\gamma$ and assuming $\tau_i=0$ for now, we obtain
\begin{eqnarray}
	N&=&\frac{2\epsilon^{2}\beta(1+2\beta)\eta}{[(\beta+1)^{2}-\beta^{2}\epsilon^{2}](1-\epsilon^{2})},\\
	M&=&\frac{2\epsilon\beta(\epsilon^{2}\beta+\beta+1)\eta}{[(\beta+1)^{2}-\beta^{2}\epsilon^{2}](1-\epsilon^{2})},
\end{eqnarray}
where $\beta=\kappa/\gamma$ is the ratio between the amplifier bandwidth and the qubit decay.
\begin{figure}
	\centering
	\includegraphics[width=\columnwidth]{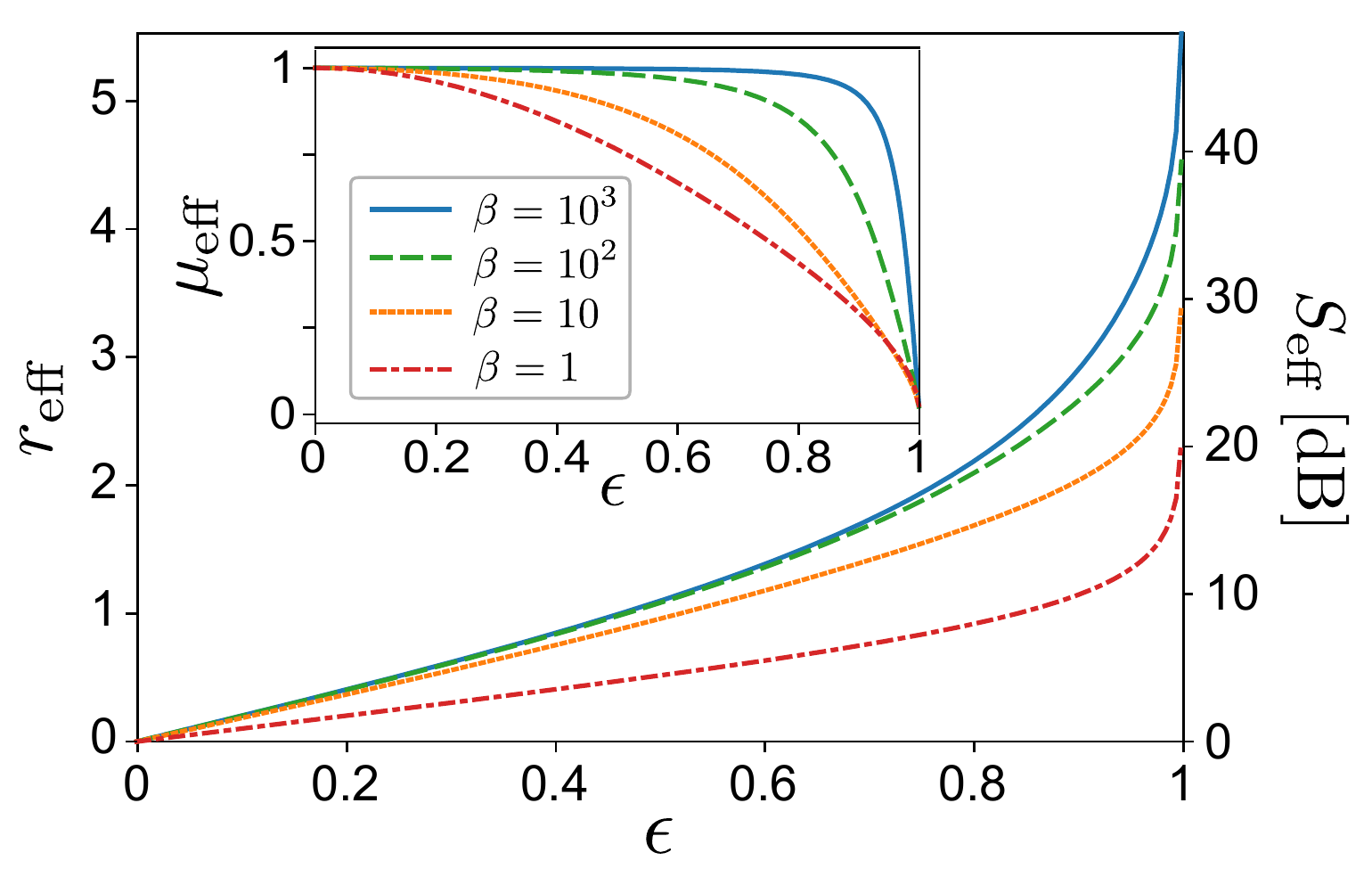}
	\caption{Dependence of the effective squeezing parameter and the effective purity on the driving strength within the FMA. The different curves are evaluated for different values of $\beta$ and assuming $\eta=1$ and symmetric conditions, $\kappa_i=\kappa$ and $\gamma_i=\gamma$. On the right axis, we plot the squeezing factor $S_{\rm eff}=e^{2r_{\rm eff}}$ in $\mathrm{dB}$. 
	Note that $r_{\rm eff}\rightarrow \infty$ and $\mu_{\rm eff}\rightarrow 0$ when $\epsilon\rightarrow1$ for all values of $\beta$.}
	\label{fig:EffParameters}
\end{figure}
As discussed in Sec.~\ref{subsec:SqueezingPurity} and shown explicitly in Fig.~\ref{fig:EffParameters}, these parameters can be reexpressed in terms of an effective squeezing parameter $r_{\rm eff}$ and an effective purity $\mu_{\rm eff}$. In this way, the fidelity of the resulting steady state can be read off directly from the general plot in Fig.~\ref{fig:MarkovContour}. 

We may also use the analytic expressions for $N$ and $M$ from above to evaluate the dependence of these quantities up to the first order corrections in the inverse bandwidth ratio $1/\beta$, 
\begin{eqnarray}
	r_{\rm eff}&\simeq& 2\tanh^{-1}(\epsilon) -\frac{1}{\beta}\frac{2\epsilon}{(1-\epsilon^2)^2},  \label{eq:BetaExpansion1} \\
	\mu_{\rm eff}&\simeq& 1-\frac{1}{\beta}\frac{4\epsilon^{2}}{(1-\epsilon^2)^2}. \label{eq:BetaExpansion2} 
\end{eqnarray}
which is valid for $\beta (1-\epsilon)^2\gg 1$.
We see that finite-bandwidth corrections get strongly amplified as one approaches the parametric instability. In particular, the purity of the effective squeezed reservoir, which is most relevant for the entanglement of the reduced qubit state, decreases significantly. Thus, even for $\beta\gg1$, it is not possible to assess the achievable amount of entanglement using a purely Markovian description.

\subsection{Optimal fidelities} 

\begin{figure}
	\centering
	\includegraphics[width=\columnwidth]{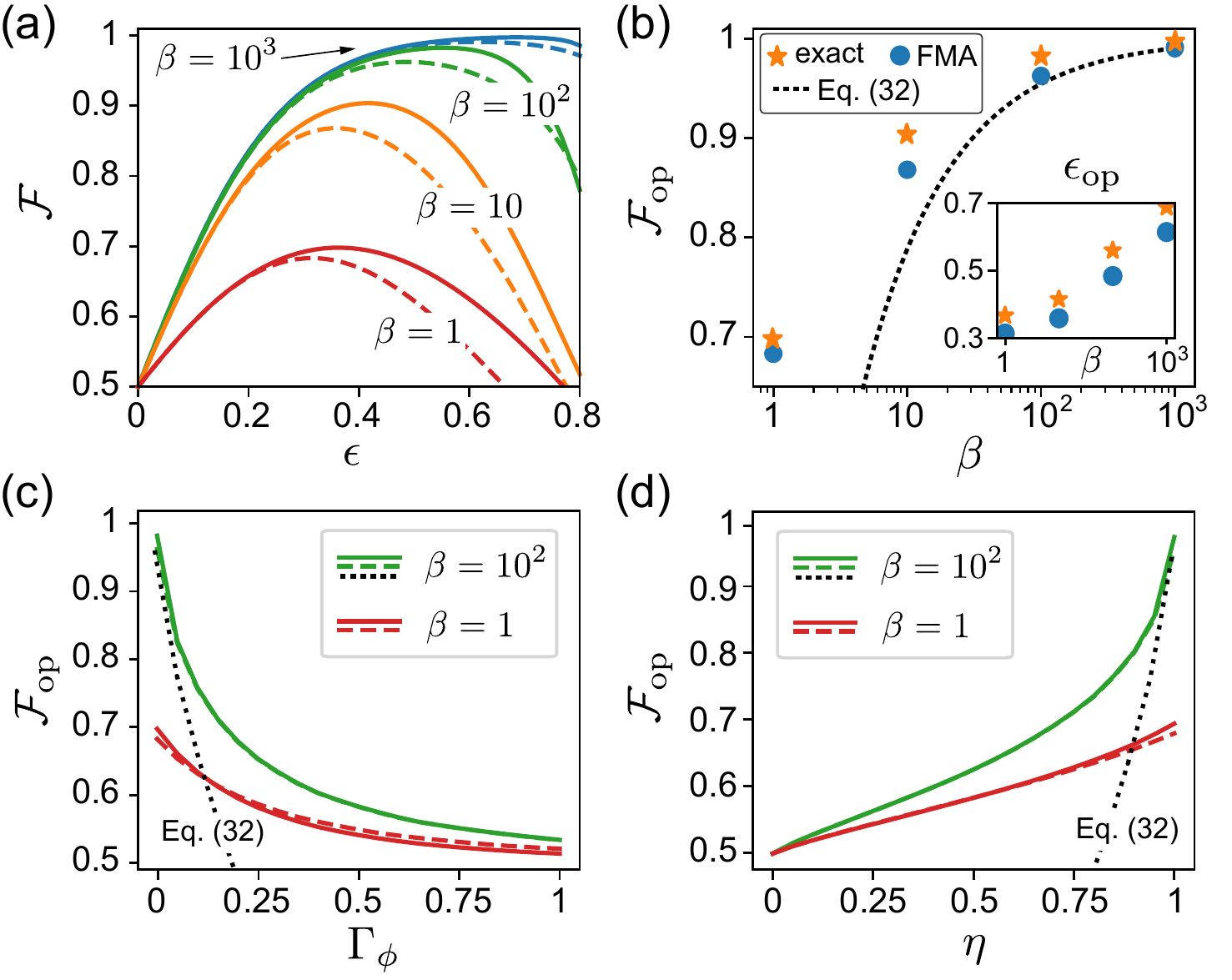}
	\caption{(a) Plot of the Bell-state fidelity $\mathcal{F}$ as a function of the driving strength and for different amplifier bandwidths, $\beta=\kappa/\gamma$. The solid lines represent the results obtained from the numerical solution of the full cascaded master equation, Eq.~\eqref{eq:CascadedME}, which are compared with the predictions under the FMA (dashed lines), assuming $\Gamma_\phi=0$ and $\eta=1$. (b) Plot of the optimal fidelity $\mathcal{F}_{\rm op}$ and the corresponding optimal driving strength $\epsilon_{\rm op}$ (inset) under the same conditions. The dotted line shows the analytic approximation given in Eq.~\eqref{eq:OptFidelity}.  (c) Plot of the optimal fidelity as a function of the qubit dephasing $\Gamma_\phi$ for $\eta=1$ and (d) as a function of the channel transmissivity $\eta$ for $\Gamma_\phi=0$. In both plots, we assume a value of $\beta=10^2$ for the upper curves and $\beta=1$ for the lower curves, where the dashed lines represent the respective FMA results. For all plots in this figure we have set $\Delta_i=0$, $\kappa_{1,2}=\kappa$ and $\gamma_{1,2}=\gamma$. }
	\label{fig:EffFidelity}
\end{figure}

Figure \ref{fig:EffFidelity} summarizes the performance of the entanglement distribution scheme in realistic settings. First of all, Fig.~\ref{fig:EffFidelity} (a) shows the dependence of the fidelity $\mathcal{F}$ on the driving strength $\epsilon$ for different ratios $\beta=\kappa/\gamma$.  Here we compare the results from a simulation of the full master equation with the predictions obtained from the FMA. In both cases, we find the expected maximum for intermediate values of $\epsilon$, which results from an increase in squeezing on the one hand and from the loss of purity  on the other hand. While for moderate and large driving strengths we see a deviation of the approximate results from exact numerics, the qualitative trends are still accurately captured. Importantly, in all the investigated parameter regimes we find that the FMA agrees with or underestimates the exact fidelity and can thus be reliably used to predict lower bounds for the achievable amount of entanglement. 

In the limit of low pump values, the FMA becomes exact and we obtain a simple analytic expression for the fidelity, 
\begin{equation}
	\mathcal{F}(\epsilon\ll 1)\simeq\frac{1}{2}+\frac{2\beta\eta}{(1+\beta)(1+2\Gamma_\phi)}\epsilon.
\end{equation}
Therefore, the fidelity increases linearly with the pump strength [see Fig.~\ref{fig:EffFidelity} (a)], with a slope that depends on all the different sources of imperfections. To estimate the maximally achievable fidelities, we assume that this maximum is reached for a pumping strength $\epsilon\approx 1$ and expand $\mathcal{F}$ to lowest order in $(1-\epsilon)$, $1/\beta$, $\Gamma_\phi$ and $(1-\eta)$,
\begin{equation}
	\mathcal{F}(\epsilon\approx 1)\simeq 1-\frac{(1-\epsilon)^4}{16} -\frac{3}{(1-\epsilon)^2}\left[\frac{1}{2\beta}+\Gamma_{\phi}+(1-\eta)\right].
\end{equation}
By optimizing this result with respect to the driving strength we obtain
\begin{equation}\label{eq:OptFidelity}
	\mathcal{F}^{\rm app}_{\rm op}\simeq 1-\frac{3\sqrt[3]{9}}{4} \left[\frac{1}{2\beta}+\Gamma_{\phi}+(1-\eta)\right]^{\frac{2}{3}}.
\end{equation}
Although the result in Eq.~\eqref{eq:OptFidelity} is based on various crude approximations, it still gives a good estimate for the overall scaling of the maximal fidelity that is achievable with this scheme in the presence of imperfections. In particular, as shown in Fig.~\ref{fig:EffFidelity} (b), for the parameter regimes of interest, $\mathcal{F}_{\rm op}>\mathcal{F}^{\rm app}_{\rm op}$, where $\mathcal{F}_{\rm op}$ is the exact optimized fidelity evaluated numerically. In Figs.~\ref{fig:EffFidelity} (c) and (d) $\mathcal{F}_{\rm op}$ is also shown for different non-ideal settings.

It is worth emphasizing that a fidelity of $\mathcal{F}>0.5$, and therefore a finite amount of steady-state entanglement, can be reached with this scheme even for rather high waveguide losses. This is in contrast to a deterministic state transfer scheme, where losses of $(1-\eta)\geq 0.5$ would not permit a finite amount of entanglement.

\subsection{The role of nonlinearities} 
All the results presented in this work are based on a linearized treatment of the parametric amplifier. For driving strengths up to $\epsilon=0.8$, the average photon number in each mode, $\langle a_i^\dag a_i\rangle\lesssim 1$, is still rather low and a linearized description should be justified in most systems of interest. However, to estimate the impact of a finite nonlinearity in the system, we can study the steady state fidelity $\mathcal{F}$ as a function of the Hilbertspace truncation number $n_{\rm trunc}$ in our numerical simulation. Such an investigation is presented in Appendix~\ref{app:Truncation} and shows that for small and moderate $\epsilon \lesssim 0.5$, the amplifier must be linear only up to photon numbers of $n_{\rm trunc}\approx 4-6$ in order to reach the same optimal fidelities as in the fully linear case. As we approach the instability, the requirements on the linearity of the amplifier become more stringent, but only when the bandwidth is sufficiently high. For example, $n_{\rm trunc}\approx 20$ is sufficient to reach $\mathcal{F}>0.98$ for $\epsilon=0.8$ and $\beta=10^3$.

\section{Steady state entanglement in asymmetric networks} 
\label{sec:LongDistance}

In our discussion so far we have neglected the finite time that it takes for the photons to propagate from the down-conversion source to the qubits. This is not a crucial assumption in situations where both qubits are located approximately equally far away from the amplifier. The photons then simply take a time $\tau_1\approx \tau_2$ to propagate from the source to the qubits, which is irrelevant for the steady state correlations. However, in situations where $\tau_1\neq \tau_2$, the qubits are driven by photons that have been emitted at two different times. When the time lag $\tau=\tau_2-\tau_1$ is too long, correlations between these photons are lost~\cite{Fedorov2018}. Such situations can occur, for example, when the amplifier and the first qubit are located in the same laboratory, while the second beam is sent via an optical fibre to a qubit at a remote location.  

\subsection{Quantum networks with propagation delays} 
For larger networks with non-negligible propagation delays, different choices for the definition of entanglement and correlations can be considered. For quantum key distribution schemes or similar applications, where the quantum states are only used once, one is typically interested in correlations between measurements that are  delayed by the respective propagation times of the transmitting photons. However, for other applications, where quantum states are redistributed within the network multiple times, the more relevant question is how much entanglement there exists between different nodes at a given point in time. In the following, we are interested in this second type of scenario, where signal delays become relevant. 

In the presence of finite propagation delays, the dynamics of the full density operator $\rho(t)$ can no longer be described by the time-local master equation in Eq.~\eqref{eq:CascadedME}, which was the basis for our exact numerical simulations so far. However, as discussed in more detail in Appendix~\ref{app:TimeDelayQLE},  this master equation can still be used to evaluate non-equal time correlation functions, which can be related to the steady state expectation values of the actual network with time delays. More precisely, given an arbitrary product of two qubit operators $O_1\equiv O_1\otimes \mathbbm{1}_2$ and  $O_2\equiv \mathbbm{1}_1\otimes O_2$, its steady state expectation value can be computed as
\begin{equation}\label{eq:NonEqualTimeCorrelation}
\langle O_1 O_2\rangle_0 = \langle O_1 (\tau_2-\tau_1) O_2\rangle_{0}\Big|_{\rm loc},
\end{equation}
assuming that $\tau_2>\tau_1$. Here, $\left. \langle O_1(\tau) O_2\rangle_{0}\right|_{\rm loc}$ denotes a non-equal time correlation function, which is evaluated with the help of the quantum regression theorem using the time-local cascaded master equation in Eq.~\eqref{eq:CascadedME}. 

Based on this relation, we express the reduced steady state of the two qubits as
\begin{equation}\label{eq:MESpins}
	\rho_q^0=\sum_{\mu,\nu=0,x,y,z}^{3}\braket{\sigma_{1}^\mu \sigma_{2}^{\nu}}_0\sigma_{1}^\mu \sigma_{2}^\nu,
\end{equation}
where $\sigma^0\equiv \mathbbm{1}$. In this way, we can employ Eq.~\eqref{eq:NonEqualTimeCorrelation} to evaluate the full two-qubit density matrix of a time-delay network through numerical simulations of the time-local master equation given in Eq.~\eqref{eq:CascadedME}. In addition,  we can again use the FMA to derive a time-local master equation for $\rho_q$ only. In this approach all the propagation delays are already included in the parameters $N_i$ and $M$ in Eq.~\eqref{eq:Eff_mode_N} and Eq.~\eqref{eq:Eff_mode_M}, which are derived from the retarded field operators in Eq.~\eqref{eq:filtered_modes}.

\subsection{Fidelities and entanglement in the presence of propagation delays} 
\begin{figure}
	\centering
	\includegraphics[width=\columnwidth]{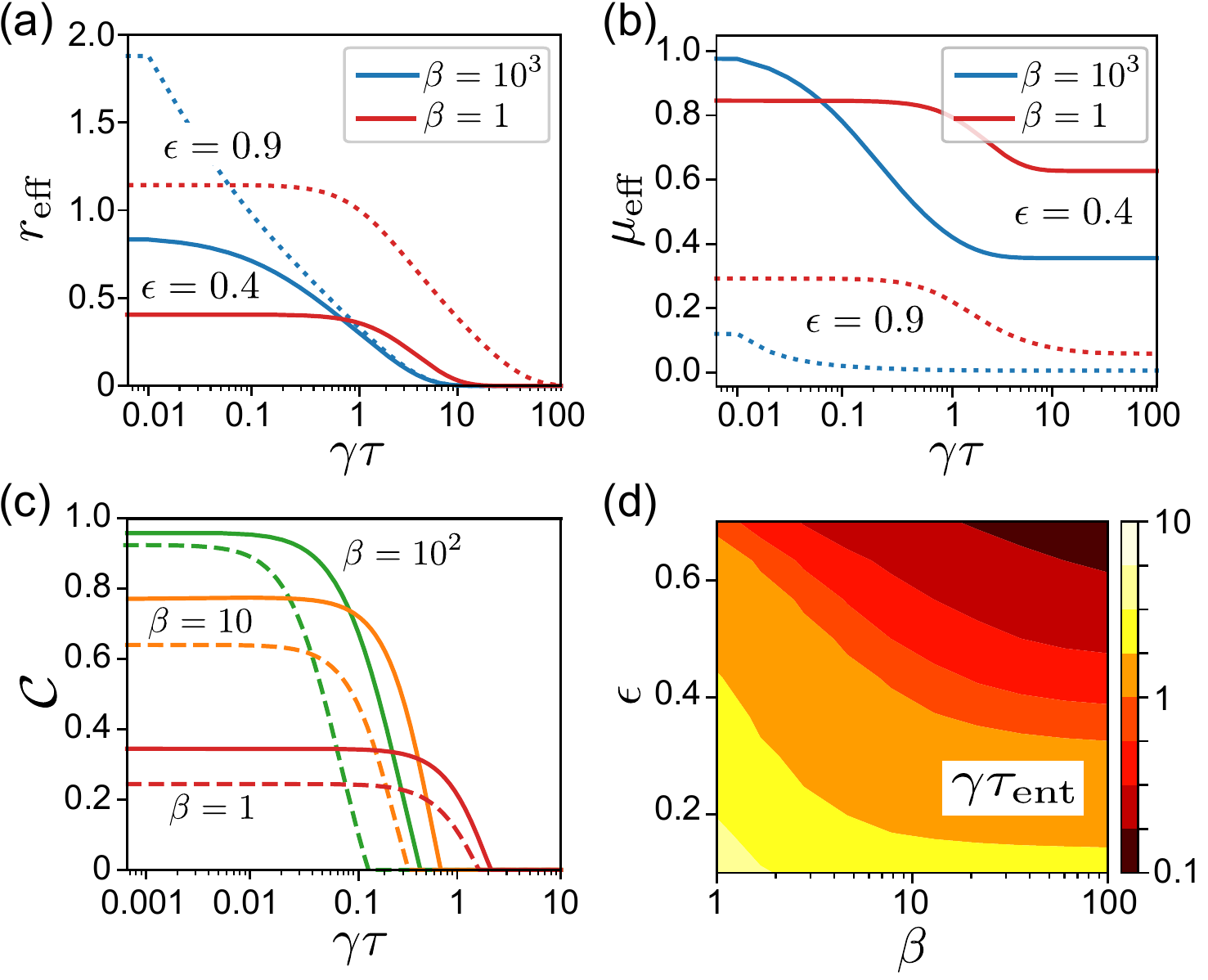}
	\caption{(a) Time delayed effective parameters $r_{\rm eff}(\tau)$ and (b) $\mu_{\rm eff}(\tau)$ for $\epsilon=0.4$ (solid) and $\epsilon=0.9$ (dotted) and for different amplifier bandwidths $\beta=1$ and $\beta=10^3$.
	(c) Steady state concurrence of a network with time delays $\mathcal{C}(\tau)$ for $\epsilon=0.5$ and different values of $\beta$. The solid lines represent the results obtained from a full numerical simulation based on Eq.~\eqref{eq:MESpins}, while the dashed lines indicate the FMA predictions. (d) Plot of the entanglement time $\tau_{\rm ent}$, i.e., the minimal delay time beyond which the entanglement vanishes, $\mathcal{C}(\tau_{\rm ent})=0$. In all plots, $\eta=1$, $\Gamma_\phi=0$ and a symmetric setup with $\Delta_i=0$ have been assumed.}
	\label{fig:TimeDelay}
\end{figure}

In Fig.~\ref{fig:TimeDelay} we consider a network with different propagation times $\tau_1\neq\tau_2$ for the two beams and study the (equal-time) steady state entanglement as a function of $\tau=\tau_2-\tau_1$. In Fig.~\ref{fig:TimeDelay} (a) and (b) we plot, first of all, the effective squeezing $r_{\rm eff}(\tau)$ and the purity $\mu_{\rm eff}(\tau)$, as obtained from the FMA for different values of $\beta$ and $\epsilon$. Within this approximation and in the limit of $\beta \gg 1$ we find that 
\begin{equation}\label{eq:MTimeDelay}
M(\tau)\simeq \frac{2\epsilon(1+\epsilon^2)}{(1-\epsilon^2)^2}e^{-\gamma\tau/2}.
\end{equation} 
This shows that in this limit the correlations decay on the timescale set by $\gamma$, and not by the relaxation rates of the amplifier, $\kappa_\pm=\kappa(1\pm\epsilon)$. Therefore, a finite amount of squeezing prevails up to delay times of about $\tau\approx \gamma^{-1}$ for all driving strengths. In contrast, the behaviour of the effective purity depends more strongly on the driving strength. This is because a larger value of $M(\tau=0)$ implies a larger absolute change of $M(\tau)$, as relevant for the purity $\mu_{\rm eff}$ defined in Eq.~\eqref{eq:mu_eff}. The timescale that determines the decay of the purity, and therefore the entanglement, can become considerably shorter than $\gamma^{-1}$ for large driving strength. This conclusion is consistent with the entanglement decay in delayed two-mode squeezed states reported in Ref.~\cite{Fedorov2018}. In the opposite regime $\beta\lesssim 1$, the squeezing and purity parameters are smaller to begin with, but they are more robust and decay only after a delay $\tau>\gamma^{-1}$, roughly independently of $\epsilon$. 

In Fig.~\ref{fig:TimeDelay} (c) we evaluate the actual steady state concurrence of this system using Eq.~\eqref{eq:MESpins} for a moderate driving strength of $\epsilon=0.5$. We see that the dependence of $\mathcal{C}(\tau)$ captures the overall trend inferred from $\mu_{\rm eff}(\tau)$. However, the exact simulations not only predict consistently higher values for $\mathcal{C}(\tau=0)$, they also show that the entanglement of the qubits is considerably more robust with respect to time delays than the entanglement of the filtered modes. In Fig.~\ref{fig:TimeDelay} (d) we define the delay time $\tau_{\rm ent}$ as the maximal delay time for which a finite amount of steady-state entanglement can still be distributed. This timescale is roughly given by $\tau_{\rm ent}\sim \gamma^{-1}$, but can be significantly reduced for very large driving strengths.

\section{Entanglement distribution rates}
\label{sec:Entanglement}

In the previous sections we have focused on the amount of entanglement that can be reached under stationary driving conditions. However, for practical applications it is equally important to know how fast this entangled state can be reached. In particular, given the possibility of distilling a highly entangled state from many copies of a state with a low amount of entanglement~\cite{Bennett1996,Duer1999,Macchiavello1999,Massar1999,Cleland2022},  it can be more favourable to optimize the generation rate rather than the fidelity.  These considerations are specifically relevant for the current entanglement distribution scheme since close to the parametric instability, where the correlations are maximized, the relaxation time of the parametric amplifier diverges. Therefore, even for an ideal, broadband amplifier, operating close to the threshold might not be the optimal choice~\cite{Govia2021}. \\

For a given application, the optimal compromise between the entanglement generation speed and the achievable fidelities depends on many details, in particular also on the local resources that are available to carry out entanglement purification protocols. To avoid such an application-specific discussion, we consider here only the following rudimentary scenario: The two-mode squeezing source is running continuously, while the qubits in each node are initialized in state $|0\rangle$. At time $t=0$ the coupling between the qubits and the photonic channels is switched on for a duration $T$, after which the qubits are decoupled again and stored in a local register. This process is then repeated with a fresh pair of qubits and so on, such that an entangled two-qubit state $\rho_q(T)$
is distributed between the two nodes every time interval $T$. 

To analyze this scenario, we consider equal qubit decay rates $\gamma_i=\gamma$ and introduce the normalized entanglement distribution rate 
\begin{equation}
\mathcal{R}=\frac{E_{F}(T)}{\gamma T}.
\end{equation}
Here $E_F(T)\equiv E_F(\rho_q(T))$ is the entanglement of formation,
which is related to the concurrence by \cite{Horodeki2009}     
\begin{equation}
	E_{F}=h\left(\frac{1+\sqrt{1-\mathcal{C}^{2}}}{2}\right),
\end{equation}
where $h(p)=-p\log_{2}(p)-(1-p)\log_{2}(1-p)$ is the Shannon entropy function. The entanglement of formation has the following meaning ~\cite{Wootters97_EntanglementFormation}. It quantifies the number of pure singlet states that are needed on average to generate the state $\rho(T)$ through LOCC operations only. This more intuitive interpretation, while still being easily computable, makes $E_{F}$ well suited for a comparative study of entanglement rates. Note, however, that $E_F$ is only an upper bound~\cite{Horodeki2009} on the number of singlet states that can be extracted from multiple copies of $\rho_q(T)$, which depends on the available purification protocols and many other details that go beyond the scope of this analysis.

\begin{figure}[t]
	\centering
	\includegraphics[width=\columnwidth]{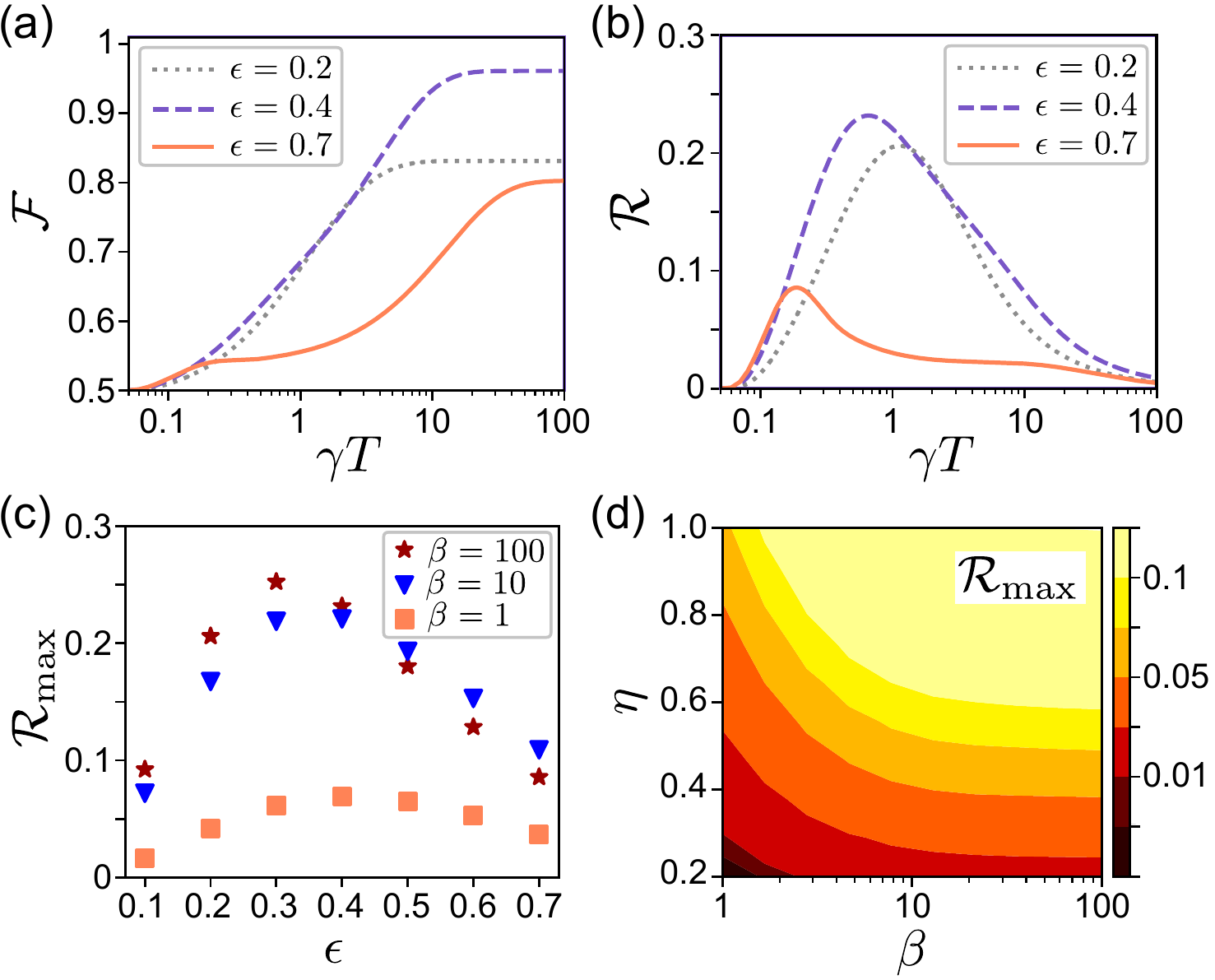}
	\caption{(a) Plot of the fidelity $\mathcal{F}$ as function of the pulse length $T$ and (b) the corresponding entanglement distribution rate $\mathcal{R}$  for different $\epsilon$. For both plots, $\beta=10$,  $\Gamma_{\phi}=0$ and $\eta=1$. (c) Dependence of the maximally achievable rate, $\mathcal{R}_{\rm max}$, on the driving strength $\epsilon$ for different values of $\beta$. (d) Plot of the entanglement rate $\mathcal{R}_{\rm max}$, optimized with respect to both $T$ and $\epsilon$, ${\rm max}_{T,\epsilon}\{ \mathcal{R}\}$, for different amplifier bandwidths and waveguide losses.}
	\label{fig:RatioOfentanglement}
\end{figure}
 
The behaviour of the entanglement distribution rate as a function of different parameters is summarized in Fig.~\ref{fig:RatioOfentanglement}.  First of all, in Fig.~\ref{fig:RatioOfentanglement} (a) we show the evolution of the fidelity $\mathcal{F}$ as a function of the interaction time $T$. While for small driving strength we find a featureless increase to the steady state value discussed in the previous sections, we observe additional small modulations for large $\epsilon$. These can be traced back to the appearance of Rabi oscillations between the states $|00\rangle$ and $|11\rangle$ in this regime.  However, these oscillations are not very pronounced and the fidelity stays well below unity in this transient regime. Therefore, while the corresponding entanglement rates shown in Fig.~\ref{fig:RatioOfentanglement} (b) reach their optimum for smaller $T$ when the driving strength is large, higher overall values for $\mathcal{R}$ can be reached for moderate $\epsilon\lesssim 0.5$.  This is shown in more detail in Fig.~\ref{fig:RatioOfentanglement} (c), where the maximal rate $\mathcal{R}_{\rm max}={\rm max}_T\{ \mathcal{R}(T)\}$ is plotted as function of $\epsilon$ for different amplifier bandwidths $\beta$. We see that the rate is maximized at values of $\epsilon\approx 0.3-0.4$. 

Finally, in Fig.~\ref{fig:RatioOfentanglement} (d) we plot to maximal entanglement rate as a function of $\beta$ and the waveguide losses $\eta$, which are the most relevant network parameters, considering otherwise ideal qubits. Again we find appreciable rates of $\mathcal{R}\gtrsim 0.1$ for a large range of parameters, including regimes with significant losses and a comparably low amplifier bandwidth. While under these conditions Bell-state fidelities close to one are not achievable, the scheme still performs very well for applications, where the distribution of a large number of weakly entangled qubits is sufficient.

\section{Discussion and Conclusions}
\label{sec:Conclusions}
In summary, we have presented a detailed analysis of a robust entanglement distribution scheme between distant qubits, which only requires moderately entangled Gaussian photonic beams. Specifically, we have investigated the performance of this scheme under realistic experimental conditions, taking a finite amplifier bandwidth, waveguide losses and also propagation delays into account. This analysis shows that while reaching extremely high fidelities of the entangled state still requires close-to-ideal conditions, the scheme is very efficient in distributing non-ideal entangled states, where it remains rather robust with respect to common experimental imperfections. For example, compared to deterministic state-transfer schemes, a finite amount of entanglement can be distributed with this scheme over very lossy channels. At the same time, the entanglement distribution rates remain high, $\mathcal{R}\sim\gamma^{-1}$, and don't suffer from the weak-driving requirement and low success probabilities of probabilistic entanglement distribution schemes. 

The analysis presented in this work was explicitly carried out for parametric amplifiers as one of the most common sources for entangled photon pairs, but it can be readily generalized for other methods~\cite{Huard2021} for producing two-mode squeezed beams. Therefore, all the results can be applied to various quantum technology platforms, where such sources are currently developed. In particular, in the field of circuit QED, highly entangled and very pure sources of entangled beams of microwave photons have been demonstrated in recent years~\cite{Eichler2011,Menzel2012,Huard2012,Devoret2016,Winkel2020,Peugeot2021}, and used for quantum illumination~\cite{Fink2020} and quantum teleportation~\cite{Fedorov2021} applications. With demonstrated purities as large as $\mu_{\rm eff}\approx0.9$ \cite{Huard2021}, such sources can be directly applied to entangle superconducting qubits over microwave channels. Similarly, in the optical domain two-mode entangled states with more than 10 dB squeezing and above-threshold purities $\mu_{\rm eff}\ge 1/\sqrt{3}$~\cite{Schnabel2007,Schnabel2013, Polzik2021} have already been demonstrated and can be further developed into efficient sources for entangling distant atoms, quantum dots or defect centers.

An important area for applications for the presented entanglement distribution scheme are hybrid quantum networks connecting qubits in the microwave and the optical domain. To entangle qubits with such vastly different frequencies, optomechanical, electro-optical and other types of quantum transducers~\cite{Han2021} are currently developed. Most of these transducers can be readily operated as a parametric amplifier by simply changing the driving frequency and be used to produce highly entangled beams of microwave and optical photons. The generation and transduction of entanglement can then be combined and optimized within a single device, which can significantly boost efficiencies and reduce the control complexity of such hybrid network architectures.

\acknowledgments
We thank Themis Mavrogordatos and Daoquan Zhu for initial contribution on the presented topic and Kirill Fedorov for stimulating discussions on entangled microwave beams. This work was supported by the Austrian Science Fund (FWF) through Grant No. P32299 (PHONED) and the European Union's Horizon 2020 research and innovation programme under grant agreement No. 899354 (SuperQuLAN). Most of the computational results presented were obtained using the CLIP cluster \cite{WebCLIP}.

\appendix
\section{Derivation of the master equation for the qubit state}
\label{sec:AppendixME}
We start by considering the cascaded master equation for qubits and the parametric amplifier, which is spelt out in Eq.~\eqref{eq:CascadedME}.
We first change to an interaction picture with respect to the free evolution, $\mathcal{L}_0 = \mathcal{L}_p + \mathcal{L}_q$, by writing $\rho(t) = e^{\mathcal{L}_0t}\rho_I(t)$. In this representation, the master equation reads
\begin{equation}\label{app_eq:ME_IA_pic}
\begin{split}
	\dot{\rho}_I(t) = \,& \mathcal{L}_{\rm casc}(t)\rho_I(t),\\
 = & \sum_{i=1,2}\sum_{s=\pm} \sqrt{\eta \gamma_i \kappa_i}  e^{-\mathcal{L}_0 t}\mathcal{K}_{i}^s e^{\mathcal{L}_0 t} \rho_I(t),
\end{split}
\end{equation}
where we defined the cascaded coupling in the interaction picture, $\mathcal{L}_{\rm casc}(t) = e^{-\mathcal{L}_0 t}\mathcal{L}_{\rm casc}e^{\mathcal{L}_0 t}$, and the superoperators $\mathcal{K}_i^+\rho=[a_i \rho,\sigma_i^+]$ and $\mathcal{K}_i^-\rho=[\sigma_i^-, \rho a^\dag_i]$. 

By assuming that the relaxation time of the amplifier is fast compared to the timescale of the qubit dynamics, we can approximate the full density operator by the product $\rho_{I}(t) \simeq\rho_{p}^0 \otimes \rho_{q,I}(t)$, where $\rho_p^0$ is the steady state of the photonic system, $\mathcal{L}_p \rho_{p}^0 =0$. Up to second order in $\mathcal{L}_{\rm casc}(t)$ and under the validity of the usual Born-Markov approximation, we can then derive a master equation for the reduced qubit state $\rho_{q,I}(t) =  \mathrm{Tr}_p\{ \rho_I(t)\}$, which is given by 
\begin{equation}
	\dot{\rho}_{q,I}(t)
	= 
	\int_{-\infty}^t\mathrm{d}\tau\,\mathrm{Tr}_p\{ \mathcal{L}_{\rm casc}(t)\mathcal{L}_{\rm casc}(\tau) \rho_{p}^0 \otimes \rho_{q,I}(t)\}.
\end{equation}
After inserting the explicit expression for $\mathcal{L}_{\rm casc}(t)$ and undoing the transformation to the interaction picture, the result can be written in a compact form as
\begin{equation}\label{eq:AppME1}
	\begin{split}
		\dot{\rho}_{q}(t)
		= &  \mathcal{L}_{q}\rho_q(t)
+ 
		\sum_{i,j=1}^2 
		\eta\sqrt{\gamma_i \gamma_j}
		\sum_{s,s^\prime=\pm} 
		\int_{-\infty}^t\mathrm{d}\tau\,C_{ij}^{ss^\prime}(t-\tau)  \\
		& 
		\times s s^\prime [\sigma^{-s}_i, e^{\mathcal{L}_q (t-\tau)} [\sigma_j^{-s^\prime},e^{-\mathcal{L}_q (t-\tau)} \rho_{q}(t)]].
	\end{split}
\end{equation}
Here we have introduced the bosonic correlation functions 
\begin{equation}\label{app_eq:Cij}
	C_{ij}^{ss^\prime}(\tau) = \sqrt{\kappa_i\kappa_j} \langle :a_i^s(\tau) a_j^{s^\prime}(0):\rangle,
\end{equation}
where we identified $a_i^{+} \equiv  a_i^\dagger$ and $a_i^{-} \equiv a_i$ and assumed the normal ordering prescription $\langle :a^s(\tau) a(0):\rangle=\mathrm{Tr}_p\{a^s e^{\mathcal{L}_p \tau}(a\rho_{p}^0)\}=\langle a^s(\tau) a(0)\rangle$, while $\langle :a^s(\tau) a^\dag (0):\rangle = \mathrm{Tr}_p\{a^s e^{\mathcal{L}_p\tau}(\rho_{p}^0 a^\dagger)\}=\langle  a^\dag(0) a^s(\tau)\rangle$.

Consistent with the Markov approximation, we also neglect the slow dynamics of the qubits in Eq.~\eqref{eq:AppME1},
which leaves us with 
\begin{equation}\label{eq:AppME2}
	\begin{split}
		\dot{\rho}_{q}(t)
		=   
		\mathcal{L}_q\rho_q(t)+ &
		\sum_{i,j=1}^2 
		\eta\sqrt{\gamma_i \gamma_j} \\
		\times& \sum_{s,s^\prime=\pm}  s s^\prime I_{ij}^{ss^\prime} (0)
		[\sigma^{-s}_i,  [\sigma_j^{-s^\prime}, \rho_{q}(t)]],
	\end{split}
\end{equation}
where $ I_{ij}^{ss^\prime}(\omega)= \int_0^\infty \mathrm{d}\tau\, C_{ij}^{ss^\prime}(\tau) e^{-i\omega\tau}$ is evaluated at resonance $\omega=0$.   To proceed, we make use of the general relation $ I_{ij}^{s,s} = (I_{ij}^{-s,-s})^*$ and additional simplifications $I_{ii}^{s,s}=I_{12}^{s,-s}=I_{21}^{s,-s} = 0$, which we derive in Appendix~\ref{sec:AppendixCorrelations} for the setup at hand. With these relations and identifying $N_i = 2\eta \mathrm{Re}\{I_{ii}^{+-}(0)\}$ and $M = \eta (I_{12}^{--}(0)+I_{21}^{--}(0))$, we end up with the master equation in Eq.~\eqref{eq:EffectiveME} together with a small Hamiltonian correction 
\begin{equation}
H^\prime= -\eta \sum_i \gamma_i {\rm Im}\{ I_{ii}^{+-}(0)\} \sigma_i^z. 
\end{equation} 
These frequency shifts can be compensated by local qubit detunings and we also find that ${\rm Im}\{ I_{ii}^{+-}(0)\}=0$ for the symmetric parameters considered in most of our examples. Therefore, this corrections is neglected in the main part of the paper.

\section{Correlation Functions and Spectra of the Parametric Amplifier}
\label{sec:AppendixCorrelations}
In this appendix, we summarize the general results for the spectra of the two amplifier modes for arbitrary parameters. Since in the operation regime of interest the amplifier is linear, this can be done most conveniently by converting the master equation in Eq.~\eqref{eq:MEParam} into an equivalent set of quantum Langevin equations~\cite{Textbooks_ZollerNoise,Collett1984} for the Heisenberg operators $a_1(t)$ and $a_2(t)$ (see also Appendix~\ref{app:TimeDelayQLE}). These equations read
\begin{eqnarray}
\dot a_1= -\left(i\Delta_1 +\frac{\kappa_1}{2}\right) a_1 + \frac{\sqrt{\kappa_1\kappa_2}\epsilon}{2} a_2^\dag -\sqrt{\kappa_1}  f_{{\rm in},1},\\
\dot a_2= -\left(i\Delta_2 +\frac{\kappa_2}{2}\right) a_2 +\frac{\sqrt{\kappa_1\kappa_2}\epsilon}{2} a_1^\dag -\sqrt{\kappa_2} f_{{\rm in},2},
\end{eqnarray} 
where $f_{{\rm in},i}$ are independent white noise operators satisfying $[f_{{\rm in},i}(t),f^\dag_{{\rm in},j}(t')]=\delta_{ij}\delta(t-t')$. By introducing the vectors $\boldsymbol{v}=(a_1,a_2^\dagger,a_2,a_1^\dagger)^\top$ and $\boldsymbol{f}=(\sqrt{\kappa_1} f_{{\rm in},1}, \sqrt{\kappa_2}f_{{\rm in},2}^\dagger, \sqrt{\kappa_2}f_{{\rm in},2}, \sqrt{\kappa_1 }f_{{\rm in},1}^\dagger)^\top$, these equations can be written in a compact form as
\begin{equation}\label{eq:QLEVector}
\dot{\boldsymbol{v}}=\mathcal{M}\boldsymbol{v}-\boldsymbol{f},
\end{equation}
where the matrix $\mathcal{M}$ is given by 
  \begin{equation}
  	\mathcal{M}=
  	\begin{pmatrix}
  		-i\Delta_1 -\frac{\kappa_1}{2}& \frac{\epsilon\sqrt{\kappa_1\kappa_2}}{2} & 0 &0 \\
  		\frac{\epsilon\sqrt{\kappa_1\kappa_2}}{2} & i\Delta_2 -\frac{\kappa_2}{2}  & 0 &0 \\
		0& 0& -i\Delta_2 -\frac{\kappa_2}{2}& \frac{\epsilon\sqrt{\kappa_1\kappa_2}}{2}  \\
  		0& 0& \frac{\epsilon\sqrt{\kappa_1\kappa_2}}{2}& i\Delta_1 -\frac{\kappa_1}{2}  \\
  	\end{pmatrix}.
  \end{equation}
 For long times, $t\rightarrow \infty$, the formal solution of Eq.~\eqref{eq:QLEVector} is 
\begin{equation}
\boldsymbol{v}(t)= - \int_{-\infty}^t d\tau\, e^{\mathcal{M}(t-\tau)} \boldsymbol{f}(\tau).
\end{equation}  
From this result we obtain the full covariance matrix in steady state, $\mathcal{V}_0=\langle \boldsymbol{v}\boldsymbol{v}^\dag\rangle (t\rightarrow \infty)$, as 
\begin{equation}
\mathcal{V}_0= \int_{0}^\infty ds\, e^{\mathcal{M}s} \mathcal{R} e^{\mathcal{M}^\dag s},
\end{equation}
where $\mathcal{R}={\rm diag}(\kappa_1,0,\kappa_2,0)$ is a diagonal matrix. Since $\mathcal{M}$ is block-diagonal, the matrix exponential and therefore the individual entries of the covariance matrix can be solved analytically. We obtain
\begin{eqnarray}
\langle a^\dag_i a_i\rangle=\frac{(\bar \kappa -\kappa_i) \bar \kappa \epsilon^2 }{4\bar \Delta^2+(1-\epsilon^2)\bar \kappa^2},\\
\langle a_1 a_2\rangle =\frac{\sqrt{\kappa_1\kappa_2}(-2i\bar \Delta+\bar \kappa)\epsilon }{4\bar \Delta^2+(1-\epsilon^2)\bar \kappa^2},
\end{eqnarray}
where $\bar \kappa=\kappa_1+\kappa_2$ and $\bar \Delta= \Delta_1+\Delta_2$. All other expectation values vanish, i.e., $\langle a_1^\dag a_2\rangle=\langle a_i^2\rangle=0$. 

To evaluate the spectra we make use of the quantum regression theorem 
\begin{equation}
\partial_\tau \langle \boldsymbol{v}(\tau) \boldsymbol{v}^\dag\rangle = \mathcal{M}  \langle \boldsymbol{v}(\tau) \boldsymbol{v}^\dag\rangle,
\end{equation}
and define 
\begin{equation}
\begin{split}
\mathcal{I}(\omega)=& \int_0^\infty d\tau \,  \langle \boldsymbol{v}(\tau) \boldsymbol{v}^\dag\rangle e^{-i\omega \tau}=  \frac{1}{i\omega\mathbbm{1}-\mathcal{M}} \mathcal{V}_0.
\end{split}
\end{equation} 
The spectra for the photon occupation numbers and correlations can be obtained elementwise as $I_{a_1^\dag a_1}(\omega)=\kappa_{1} [\mathcal{I}(\omega)]_{44}$, $I_{a_2^\dag a_2}(\omega)=\kappa_{2} [\mathcal{I}(\omega)]_{22}$, $I_{a_1 a_2}(\omega)=\sqrt{\kappa_{1}\kappa_{2}} [\mathcal{I}(\omega)]_{12}$ and $I_{a_2 a_1}(\omega)=\sqrt{\kappa_{1}\kappa_{2}} [\mathcal{I}(\omega)]_{34}$. The general expressions for the spectra are rather lengthy, but for the symmetric case, $\kappa_i=\kappa$ and $\Delta_i=\Delta$, we obtain
\begin{multline}
	I_{a^{\dagger}_i a_i}(\omega)=\frac{2\epsilon^2\kappa^3(\kappa+i\omega)}{(\kappa^2-\delta^2)(\kappa^2-\delta^2+4i\kappa\omega-4\omega^2)},
\end{multline}
and
\begin{multline}
	I_{a_{1}a_{2}}(\omega)=\frac{\epsilon\kappa^2(\delta^2+\kappa^2-4i\Delta(\kappa+i\omega)+2i\omega\kappa)}{(\kappa^2-\delta^2)(\kappa^2-\delta^2+4i\kappa\omega-4\omega^2)},
\end{multline}
where  $\delta^2=\epsilon^2\kappa^2-4\Delta^2$. These results reduce to the expressions in Eq.~\eqref{eq:MarkovN} and Eq.~\eqref{eq:MarkovM} for $\Delta=0$. Note that for a symmetric amplifier ${\rm Im}\{ I_{a^{\dagger}_i a_i}(0)\}=0$ and there is no Lamb-shift contribution. In general this is not the case.

\section{Hilbertspace truncation}\label{app:Truncation}
\begin{figure}
	\centering
	\includegraphics[width=\columnwidth]{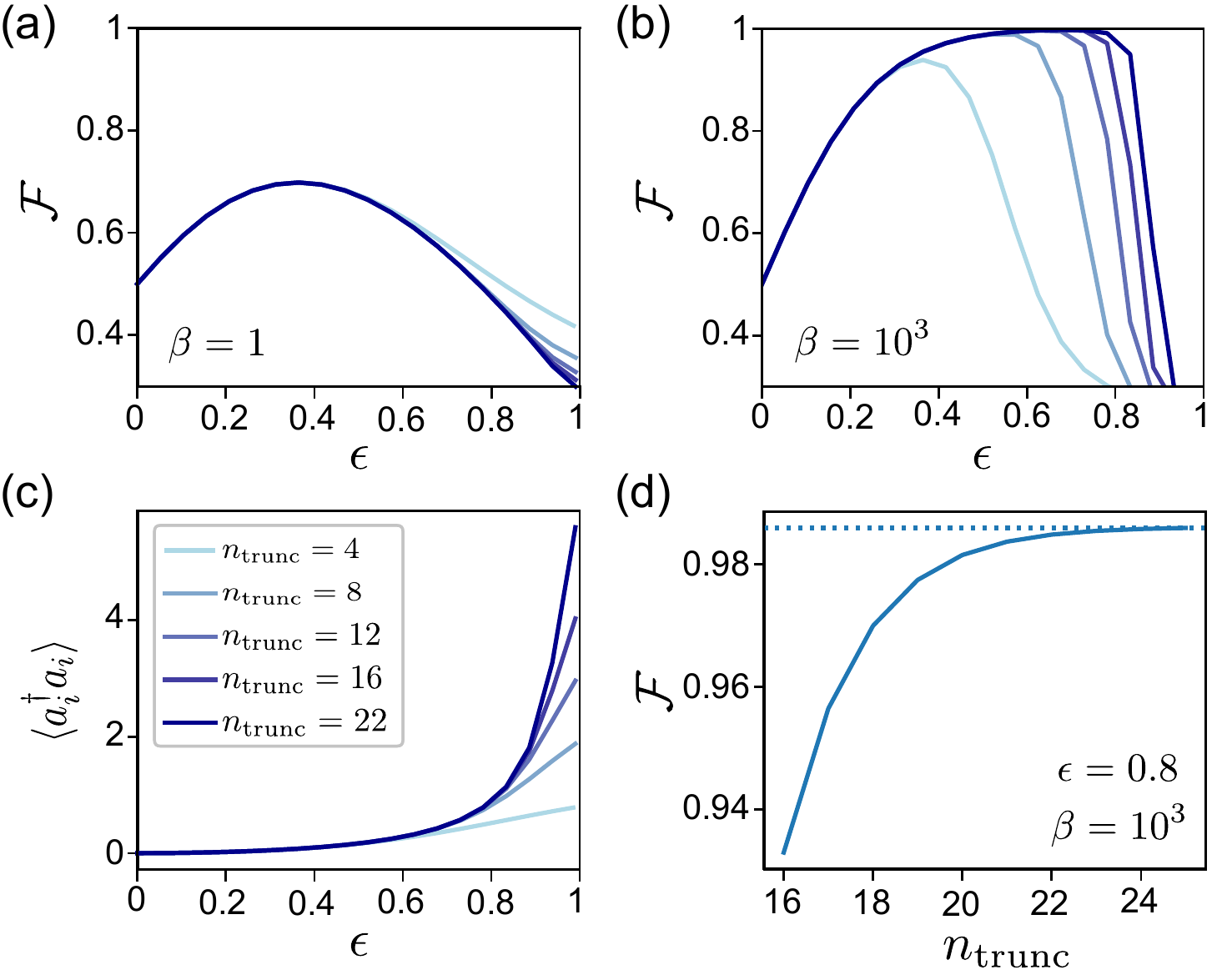}
	\caption{Dependence of the numerically evaluated fidelity $\mathcal{F}$ on the photon truncation number $n_{\rm trunc}$ for (a) $\beta=1$ and (b) $\beta=10^3$. (c) Plot of the average number of photons in each of the parametric amplifier modes as a function of the driving strength and for different $n_{\rm trunc}$. (d) Convergence of the fidelity $\mathcal{F}$ with increasing photon truncation number for the case of $\beta=10^3$ at $\epsilon=0.8$. For all plots, $\Delta_i=0$, $\eta=1$, $\Gamma_\phi=0$ and otherwise symmetric conditions have been assumed. }
	\label{fig:Truncation}
\end{figure}

In our numerical simulations of the full cascaded master equation we truncate the Hilbertspace of each photon mode and only include a finite set of number states $|n\rangle$ with $n< n_{\rm trunc}$. In Fig.~\ref{fig:Truncation} (a) and (b) we plot the entanglement fidelity for different truncation numbers $n_{\rm trunc}$ and two very different values of $\beta$. Fig.~\ref{fig:Truncation} (c) shows the corresponding average photon number in each of the amplifier modes. We see that even though for driving strengths of $\epsilon \lesssim 0.8$ the average number of photons remains small, $\langle a^\dag_ia_i\rangle <1$, in the case of large $\beta$ a substantially larger $n_{\rm trunc}$ must be used in order to capture the qubit-qubit entanglement accurately. For even larger $\epsilon$ the average photon number grows considerably and exact simulations of the full master equation become very demanding. For this reasons all our simulations are restricted to the parameter regime $\epsilon\leq 0.8$ and $\beta\leq 10^3$, where a convergence to the required level of accuracy can be achieved with $n_{\rm trunc}=25$ [see Fig. ~\ref{fig:Truncation} (d)].

\section{Quantum Langevin equations for cascaded networks with propagation delays}\label{app:TimeDelayQLE}
Instead of working with a master equation for the system density operator $\rho(t)$, an equivalent description of open quantum systems can be obtained from a set of quantum Langevin equations for Heisenberg operators~\cite{Textbooks_ZollerNoise}. For the considered setup, these quantum Langevin equations can be derived from a unidirectional system-waveguide interaction of the form
\begin{equation}
H_{\rm int}= i \sum_{i=1,2} \left[ \sqrt{\kappa_i} F^\dag_i(0)  a_i + \sqrt{\gamma_i} F^\dag_i(d_i)  \sigma_i^-  - {\rm H.c.}\right],
\end{equation}
where 
\begin{equation}
F_i(z)= \frac{1}{\sqrt{2\pi}} \int_0^\infty  d\omega\,  e^{i\omega z/c_i} b_i(\omega)
\end{equation}
is the field operator for the $i$-th channel at position $z$ and $[b_i(\omega),b^\dag_j(\omega')]=\delta_{ij}\delta(\omega-\omega')$.  The equation of motion for an arbitrary system operator $O(t)$ then obeys
\begin{equation}\label{eq:QLE}
\begin{split}
\dot O=  &\, i[H_{\rm sys},O] - \sum_{i=1,2}  \left(  \sqrt{\kappa_i} [O,a_i^\dag] F_i(0,t)  + {\rm H.c.} \right) \\
&- \sum_{i=1,2}  \left(  \sqrt{\gamma_i} [O,\sigma_i^+]  F_i(d_i,t)  + {\rm H.c.} \right),
\end{split}
\end{equation}
where $H_{\rm sys}$ is the bare system Hamiltonian. In turn, under the validity of the usual Born-Markov approximation, the field operator is given by~\cite{Textbooks_ZollerNoise} 
\begin{equation}\label{eq:QLEField}
\begin{split}
F_i(z,t)=& \, F_{{\rm in},i}(z,t) +  \sqrt{\kappa_i} \Theta(z)a_i(t-z/c_i) \\
&+ \sqrt{\gamma_i} \Theta(z-d_i)\sigma_i^-(t-z/c_i+\tau_i),
\end{split}
\end{equation}
where we have omitted nonessential propagation phases and $\Theta(x)$ denotes the unit step function. The field $F_{{\rm in},i}(z,t)$ is the free field operator in the $i$-th channel and obeys 
\begin{equation}\label{eq:FinPropagation}
F_{{\rm in},i}(z,t+\tau )= F_{{\rm in},i}(z-c_i\tau,t).
\end{equation} 
Since we neglect thermal excitations, we find that $F_{{\rm in},i}(z,t)\rho_{\rm full}^0=\rho_{\rm full}^0 F^\dag_{{\rm in},i}(z,t)=0$, where  $\rho_{\rm full}^0$ is the initial state of the full network. We identify $f_{{\rm in},i}(t) = F_{{\rm in},i}(z=0,t)$ with the noise operators used in Appendix~\ref{sec:AppendixCorrelations}.

Note that for every position $z$, the fields $F_i(z,t)$ represent independent degrees of freedom and thus
\begin{eqnarray}
[F_i(z,t), O(t)] &=& 0, \\
\left[ F_i(z,t), F_j(z',t) \right] &=& c_i \delta_{ij} \delta(z-z').
\end{eqnarray}
Together with Eq.~\eqref{eq:QLEField} and Eq.~\eqref{eq:FinPropagation}, these relations can be use to derive all the non-equal time correlation functions used below.

\subsection{Weak driving limit} 
From Eq.~\eqref{eq:QLE} we can readily derive expectation values of qubit observables in the weak excitation limit, $\epsilon\ll 1$. In this limit we can approximate $\sigma^z_i(t)\simeq -1$ and obtain
\begin{equation}
\begin{split} 
\dot \sigma_i^-(t) \simeq  -\frac{\gamma_i}{2} \sigma_i^-(t)  - \sqrt{\gamma_i \eta} f_{{\rm out},i}(t-\tau_i) \\
-\sqrt{\gamma_i (1-\eta)} f_{{\rm in},i}^\prime(t).
\end{split} 
\end{equation} 
Here $f_{{\rm out},i}(t)\equiv F_{{\rm out},i}(z=0,t)$, where $F_{{\rm out},i}(z,t)=F_{{\rm in},i}(z,t)+\sqrt{\kappa_i} a_i(t-z/c_i)$, and $f_{{\rm in},i}^\prime(t)$ is an independent noise operator, which we have included to account for waveguide losses.  In the limit $t\rightarrow \infty$ we obtain
\begin{equation}
\sigma_i^-(t) = -\sqrt{\gamma_i  \eta } \int_{-\infty}^t {\rm d} s \, e^{-\gamma_i(t-s)/2}  f_{{\rm out},i}(s-\tau_i), 
\end{equation} 
where contributions from $f_{{\rm in},i}^\prime(t)$, which always act on the vacuum state, have already been omitted. For the evaluation of the expectation value $\langle \sigma_i^+\sigma_i^-\rangle(t\rightarrow \infty)=N_i$ we use 
\begin{equation}
\langle f^\dag_{{\rm out},i}(t) f_{{\rm out},i}(t')\rangle=\kappa_i \langle a^\dag_i(t)a_i(t')\rangle
\end{equation}
and after some manipulations we obtain the result for $N_i$ given in Eq.~\eqref{eq:Eff_mode_N}. For the evaluation of the correlations, $\langle \sigma_1^-\sigma_2^-\rangle(t\rightarrow \infty)=M$, we must take into account that $f_{{\rm in},1}(t)$ and $a_2(t')$ do not commute in general and \cite{Collett1985}
\begin{equation}
\langle f_{{\rm out},1}(t) f_{{\rm out},2}(t')\rangle=\sqrt{\kappa_1\kappa_2}  \langle   \mathcal{T} a_1(t)a_2(t')\rangle,
\end{equation}
where $\mathcal{T}$ denotes the time-ordering operator. Again we find that the resulting expression for $M$ matches Eq.~\eqref{eq:Eff_mode_M}. Therefore, this comparison shows that the results obtained within the FMA  for the steady state of the qubits become exact in the weak driving limit, $N_i, |M|\ll1$. 

\subsection{Propagation delays}  
To handle finite propagation delays in a more general manner, we introduce a set of shifted Heisenberg operators
\begin{equation}
\tilde \sigma_i^k(t)=\sigma_i^k(t+\tau_i),\qquad \tilde a_i(t)=a_i(t+\tau_\varepsilon).
\end{equation}
Here, $\tau_\varepsilon$ is a small time delay, which is negligible on the timescale of the system dynamics, but must be kept finite when evaluating commutation relations with the waveguide fields. 

The shifted Heisenberg operators obey (assuming $\eta=1$ for simplicity) 
\begin{eqnarray}\label{eq:tildedot1}
\dot{\tilde a}_i(t) &=  &i [\tilde H_p(t), \tilde a_i(t)] -\frac{\kappa_i}{2}\tilde a_i(t)  \nonumber \\
&&- \sqrt{\kappa_i} f_{{\rm in},i}(t+\tau_\varepsilon), \\
\dot{\tilde\sigma}_i^-(t) &=  &-\frac{\gamma_i}{2}\tilde \sigma_i^-(t) +  \sqrt{\gamma_i \kappa_i} \tilde\sigma_i^z(t)  a_i(t)  \nonumber\\
&&+ \sqrt{\gamma_i} \tilde \sigma_i^z(t) f_{{\rm in},i}(t), \\
\dot{\tilde\sigma}_i^z(t) &=  &-2\gamma_i \tilde \sigma_i^+(t)\tilde \sigma_i^-(t)  \nonumber \\
&&-  2\sqrt{\gamma_i \kappa_i}  \left[\tilde\sigma_i^+(t) a_i(t) +  a^\dag _i(t) \tilde\sigma_i^-(t)\right]  \nonumber \\
&&- 2\sqrt{\gamma_i} \left[\tilde\sigma_i^+(t) f_{{\rm in},i}(t) + f^\dag_{{\rm in},i}(t)\tilde\sigma_i^-(t)\right]. \label{eq:tildedot3}
\end{eqnarray}
Since in these equations all $f_{{\rm in},i}(t)$ operators appear to the right and all $f^\dag_{{\rm in},i}(t)$ operators to the left,  we can perform the expectation value with respect to the initial vacuum state $\rho_{\rm full}^0$ and take the limit $\tau_\epsilon\rightarrow 0$ afterwards. As a result, the expectation values for $\langle \dot{\tilde\sigma}_i^k(t) \rangle$ and $\langle \dot{\tilde a}_i(t) \rangle$ do not explicitly depend on the delay times $\tau_i$ anymore and their expressions are identical to the ones obtained for $\langle \dot{\sigma}_i^k(t) \rangle$ and $\langle \dot{ a}_i(t) \rangle$ from the cascaded master equation given in Eq.~\eqref{eq:CascadedME}.

As a next step we show that the same is true for arbitrary operator products $\tilde S_1(t)\tilde S_2(t) \tilde A(t)$, where $\tilde S_i(t)$ are Pauli operators and $\tilde A(t)$ is an arbitrary product of operators $\tilde a_{1,2}(t)$ and  $\tilde a^\dag_{1,2}(t)$. To evaluate the time derivative of this product, we apply the product rule and use the time derivatives for the individual operators given in Eqs.~\eqref{eq:tildedot1}-\eqref{eq:tildedot3}. This results in terms of the form 
\begin{equation}\label{eq:ProductRule}
\begin{split} 
&\frac{d}{dt}  \tilde S_1(t)\tilde S_2(t) \tilde A(t)= \dots \\
&-\sqrt{\gamma_1} [ \tilde S_1(t),\tilde \sigma_1^+(t)] F_1(d_1,t+\tau_1) \tilde S_2(t) \tilde A(t) +\dots \\
&- \sqrt{\gamma_2} \tilde S_1 (t) F_2^\dag (d_2,t+\tau_2) [ \tilde \sigma_2^-(t), \tilde S_2(t)]  \tilde A(t) + \dots \\
&-\sqrt{\kappa_1} \tilde S_1(t)\tilde S_2(t) F_1^\dag(0,t+\tau_\epsilon) [ \tilde a_1(t), \tilde A(t)]+\dots 
\end{split}
\end{equation}
Before we can take the expectation value with respect to $\rho_{\rm full}^0$, all $f_{{\rm in},i}(t)$ operators must be commuted to the right and all $f^\dag_{{\rm in},i}(t)$ operators to the left. To do so we write, for example, 
\begin{equation}
F_1(d_1,t+\tau_1)=\frac{\sqrt{\gamma_1}}{2} \tilde \sigma_1^-(t)+ F_{{\rm out},1}(0,t),
\end{equation}
and use $F_{{\rm out},1}(0,t)=F_1(c\tau_\epsilon,t+\tau_\epsilon)$ to show that $[F_{{\rm out},1}(0,t),\tilde A(t)]=0$. Further, we write $\tilde S_2(t)=S_2(t)+\Delta S_2(t)$, where $\Delta S_2(t)$ depends on the field $F_{{\rm out},2}(d_2,t')$ for times $t'\in [t,t+\tau_2]$ only. Equivalently, it depends on the field $F_{2}(z,t)$ located in the region $z\in (0, d_2]$ and the operator 
\begin{equation}
F_{{\rm out},2}(0,t)= F_{2}(0,t)+\frac{\sqrt{\kappa_2}}{2}a_2(t).
\end{equation}
This implies that also $[\tilde S_2(t), F_{{\rm out},1}(0,t)]=0$ and $F_{{\rm out},1}(0,t)=f_{{\rm in},1}(t)+\sqrt{\kappa_1}a_1(t)$ in the second line of Eq.~\eqref{eq:ProductRule} can be commuted all the way to the right. Similar arguments can also be made for all the other terms to achieve the desired operator ordering.

In summary, from this derivation we obtain a set of coupled equations of motion for the expectation values of arbitrary operator products $\langle \tilde S_1(t)\tilde S_2(t) \tilde A(t)\rangle$, which are independent of the delay times $\tau_i$ and have the same structure as the corresponding equations of motion derived from the time-local cascaded master equation. After taking again the limit $\tau_\epsilon\rightarrow 0$, this result implies that
\begin{equation}\label{eq:ExpectationValueRelation}
\langle S_1(t+\tau_1) S_2(t+\tau_2)A(t) \rangle = \langle S_1 S_2 A\rangle(t)\Big|_{\rm loc},
\end{equation}
assuming appropriately matched initial conditions.

In the main text we are interested in equal-time expectation values $\lim_{t\rightarrow\infty} \langle S_1(t) S_2(t)\rangle$, for which Eq.~\eqref{eq:ExpectationValueRelation} can not be directly applied. Instead we repeat the whole derivation for the operators $\tilde S_1(t+t_0)$ and take the average with respect to the state $S_2(t_0+\tau_2)\rho_{\rm full}(t_0)$. Since $S_2(t_0+\tau_2)$ depends on $f_{{\rm in},i}(t)$ for $t\leq t_0$ only, it commutes with the relevant noise terms $\sim f_{{\rm in},i}(t+t_0)$ and we can still make use of $f_{{\rm in},i}(t+t_0)S_2(t_0+\tau_2)\rho_{\rm full}(t_0)=0$. Therefore, this approach provides us with the relation 

\begin{eqnarray}
\langle S_1(t+\tau_1+t_0) S_2(t_0+\tau_2) \rangle & =&  \\
\langle S_1(t+t_0) S_2(t_0) \rangle \Big |_{\rm loc}, \nonumber
\end{eqnarray}
which extends the result from above to more general correlations. By assuming that $t_0$ is long enough such that the system has reached a steady state, we can redefine  $t_0\rightarrow t_0-\tau_2$ and set $t=\tau_2-\tau_1$. This leaves us with the result stated in Eq.~\eqref{eq:NonEqualTimeCorrelation}.

\end{document}